\documentclass[aps,prd,showpacs,onecolumn,preprintnumbers,notitlepage,nofootinbib,tightenlines,10pt]{revtex4-1}
\usepackage{amsmath,bm}
\usepackage{times}
\usepackage{dcolumn}
\usepackage{amsfonts}
\usepackage{amssymb}
\usepackage{braket}
\usepackage{color,graphicx}
\usepackage{epstopdf}
\usepackage{latexsym}
\usepackage[dvipsnames]{xcolor}
\usepackage{slashed}
\usepackage{CJK,upgreek,fancyhdr}
\usepackage{float}
\usepackage{multirow}
\usepackage{hyperref}
\hypersetup{colorlinks,citecolor=nicegreen,linkcolor=nicered,urlcolor=RoyalBlue}
\usepackage{enumitem}
\usepackage{natbib}
\usepackage{relsize}
\usepackage[left=2.5cm,right=2.5cm,top=2.0cm,bottom=2.5cm]{geometry}
\newcommand{\beq}{\begin{eqnarray}}
\newcommand{\eeq}{\end{eqnarray}}
\newcommand{\non}{\nonumber\\ }

\newcommand{\psl}{ P \hspace{-2.4truemm}/ }
\newcommand{\nsl}{ n \hspace{-2.2truemm}/ }
\newcommand{\vsl}{ v \hspace{-2.2truemm}/ }
\newcommand{\epsl}{\epsilon \hspace{-1.6truemm}/\,  }

\def\lsim{ {\ \lower-1.2pt\vbox{\hbox{\rlap{$<$}\lower6pt\vbox{\hbox{$\sim$}
		}}}\ } }
\def\gsim{ {\ \lower-1.2\vbox{\hbox{\rlap{$>$}\lower6pt\vbox{\hbox{$\sim$}
		}}}\ } }

\definecolor{Red}{rgb}{1.,0.,0.}
\definecolor{Blue}{rgb}{0.,0.,1.}
\definecolor{RoyalBlue}{rgb}{0.0,0.14,0.4}
\definecolor{nicered}{rgb}{0.7,0.1,0.2}
\definecolor{nicegreen}{rgb}{0.1,0.4,0.2}

\def\orcid#1{\kern .08em\href{https://orcid.org/#1}{\includegraphics[keepaspectratio,width=0.76em]{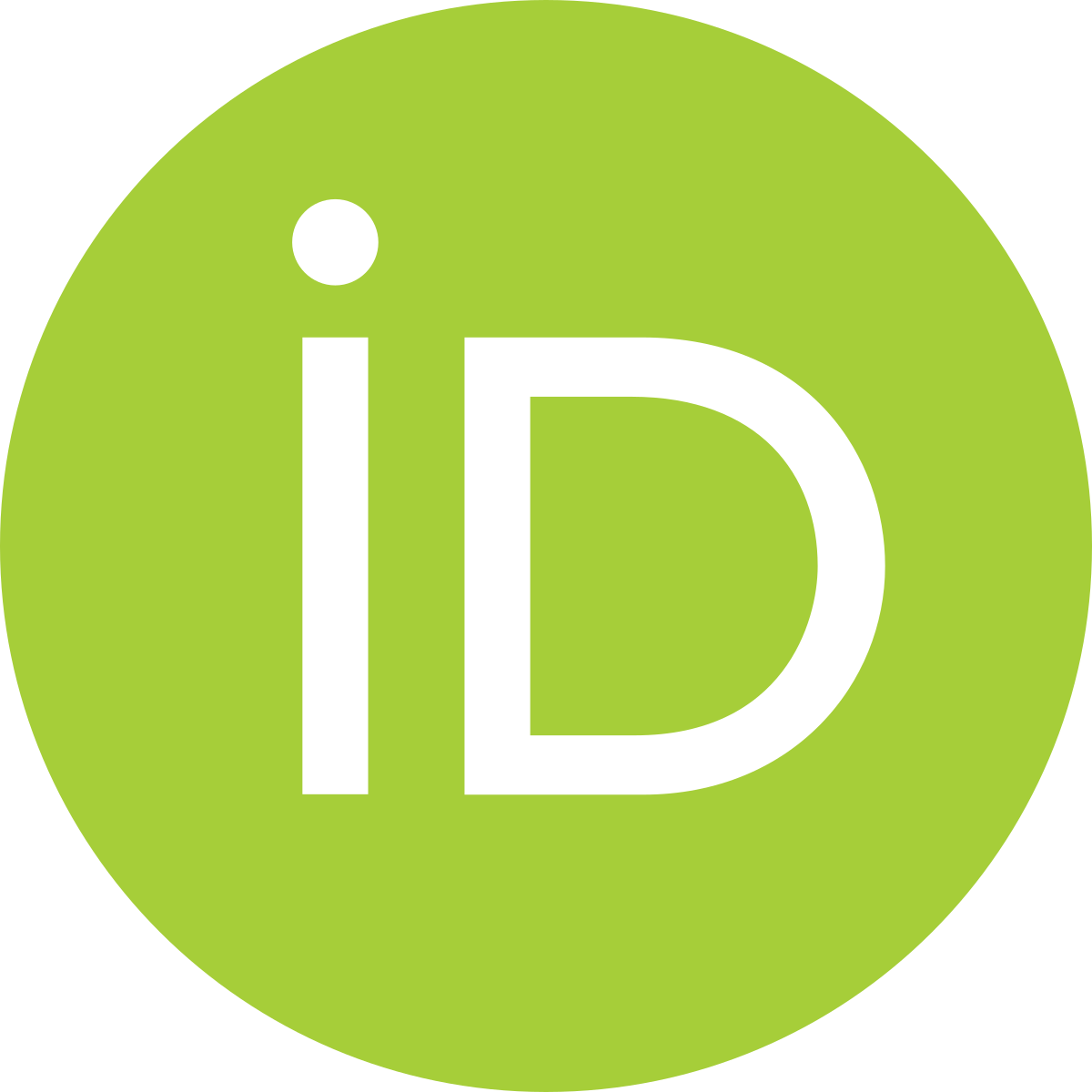}}}

\begin{document}
\title{\boldmath Pure annihilation decay $B_c^+ \to ST$
with improved perturbative QCD formalism }
\author{Sheng-Yu~Wang\orcid{0009-0001-6459-5045}}
\author{Xin~Liu\orcid{0000-0001-9419-7462}}
\email
{ liuxin@jsnu.edu.cn}
\affiliation{Department of Physics,
\href{https://ror.org/051hvcm98}{Jiangsu Normal University}, Xuzhou 221116, China}

\author{Zhen-Jun~Xiao\orcid{0000-0002-4879-209X}}
\affiliation{Department of Physics and Institute of Theoretical Physics,\\
\href{https://ror.org/036trcv74}{Nanjing Normal University}, Nanjing 210023, China}

	
\date{\today{}}
	
\begin{abstract}
We study the pure annihilation decay $B_c^+ \to ST$ for the first time in the standard model
with $S$ and $T$ representing the states of the light scalar ($1^3\!P_0$) and tensor ($1^3\!P_2$)
correspondingly. Utilizing the assignment $q\bar q$ for light scalars with two-scenarios
classification, we calculate the branching ratios of $B_c^+ \to ST$ with the improved perturbative
QCD (iPQCD) formalism at leading order in the strong coupling $\alpha_s$. The large branching
ratios around $10^{-6}$ within uncertainties in the Cabibbo-Kobayashi-Maskawa-favored channels,
such as $B_c^+ \to a_0(1450) a_2(1320)$, $B_c^+ \to a_0(980, 1450)^+ f_2(1270)$, $B_c^+ \to
K_0^*(700, 1430) K_2^*(1430)$, and $B_c^+ \to f_0(500, 1370, 1500) a_2(1320)^+$, are obtained
in the iPQCD formalism, which would be tested at the Large Hadron Collider-beauty detector in the
(near) future. The relevant tests to these theoretical predictions of large branching ratios
associated with typical relative ratios could provide important opportunities for understanding the
QCD dynamics of the involved modes or light scalars, as well as for exploring the reliability of
the iPQCD formalism. Moreover, the direct charge-parity violating asymmetries in the decay
$B_c^+ \to ST$ are naturally zero, since only tree-diagram contributions are involved.
\end{abstract}
	
\preprint{\footnotesize  JSNU-HEP-OCT.2026}
\maketitle

	
\newpage
%
%
\section{Introduction}
\label{sect:1}

The $B_c^+$ meson, which consists of two distinct heavy-flavor quarks (a bottom quark and a charm quark),
was first observed in 1998 by the CDF collaboration at the Fermilab Tevatron through its semileptonic
decay modes~\cite{CDF:1998axz,CDF:1998ihx}. Due to its unique composition, the $B_c^+$ meson exhibits a
larger phase space than its lighter $B$ partners, namely, $B^+$, $B^0$, and $B_s^0$. Since the Large Hadron
Collider (LHC) began operations in 2009, the study of $B_c^+$-meson physics has entered a new era. An extensive range of decay channels for the $B_c^+$ meson are then believed to become accessible as
experimental capabilities continue to improve in the present and the future. Therefore, comprehensive investigations of the $B_c^+$-meson decays from both theory and experiment will offer insights into
the dynamics of flavor physics.

On the experimental side, distinct from the conventional $q\bar q$ configuration predicted by the quark model, 
the particles produced in heavy-meson weak decays, e.g., the $f_0(500)$ (also known as $\sigma$)
in $D^+ \to \pi^- \pi^+ \pi^+$~\cite{E791:2000vek}, the $f_0(980)$ from $B^+ \to \pi^+ \pi^- K^+$~\cite{Belle:2002ryx},
and the $\chi_{c1}(3872)$ (also known as $X(3872)$) from $B^\pm \to \pi^+\pi^- J/\psi K^\pm$~\cite{BESIII:2023hml}, etc.,
have stimulated a unique approach for exploring exotic states~\cite{Rosner:2003ia} ever since entering the 21st century. 
Furthermore, as claimed in~\cite{Vagnoni:2025qfv}, it is very gratifying that the Large Hadron Collider-beauty (LHCb) detector
has completed its second major upgrade and resumed in 2022, which can give unprecedented and unique scientific
opportunities in flavor physics and make it the ideal laboratory for hadron spectroscopy studies. And what is more important,
``{\it it will become possible for the first time to study the production of exotic hadrons with open or hidden charm in
$B_c^+$-meson decays, extending the mass range that is currently accessible in $B^+$ decays}"~\cite{Vagnoni:2025qfv}.

On the theoretical side, rather than studying the hadron properties directly with nonperturbative Quantum ChromoDynamics
(QCD) in hadron physics, numerous studies on the productions of exotic hadrons have been performed in the $B$-meson
decays, e.g., see Refs.~\cite{Cheng:2005nb,Cheng:2007st,Cheng:2013fba,Shen:2006ms,Wang:2006ria,Dou:2015mka,Zou:2016yhb,Zou:2017iau,
Su:2019vbu,Liu:2019ymi,Liu:2021gnp,Niu:2021pcx,Li:2025ddc},
as well as in the heavier $B_c^+$-meson decays, e.g., see Refs.~\cite{Liu:2010kq,Zou:2017yxc,Liu:2023kxr,Zhang:2025yvt},
within the QCD-based factorization framework, for example,
QCD factorization approach based on collinear factorization theorem~\cite{Beneke:1999br},
perturbative QCD (PQCD) approach based on $\boldsymbol{k}_T$ factorization theorem
(with $\boldsymbol{k}_T$ being the transverse momentum of valence quark in a hadron)~\cite{Keum:2000wi,Lu:2000em},
and so on. It is well known that scalar states possess
the same quantum numbers as the vacuum, which has consequently attracted
considerable interest from the community. Notwithstanding significant efforts, the internal
structure of light scalars, particularly those light scalar states below 1~GeV, remains the subject of considerable
debate~\cite{ParticleDataGroup:2024cfk}. Currently, it is widely accepted that the light scalars
can be organized into two distinct nonets: one below
1~GeV containing $K^*_0(700)$ (also referred to as $\kappa$), $a_0(980)$, $f_0(500)$, and $f_0(980)$, and another
around 1.5~GeV including $K^*_0(1430)$, $a_0(1450)$, $f_0(1370)$ and $f_0(1500)$.
However, there is no definitive answer to the challenging question that whether they are conventional $q\bar q$ pairs,
multiquark configurations, meson-meson bound states, or hybrid admixtures. It is evident that more studies from
both theory and experiment are very important in helping to find the correct answer about the internal structure of light scalars.

To provide more evidences for shedding light on the internal structure of light scalars,
we will perform a first study in this work of the charmless hadronic $B_c^+ \to ST$ decay,
which occurs via pure annihilation, in the improved perturbative QCD (iPQCD) formalism~\cite{Liu:2018kuo,Liu:2020upy,Liu:2023kxr}.
Here, $S$ and $T$ stand for a scalar and a tensor state, respectively, consisting of light quarks.
As a sequential study following the pure annihilation nonleptonic decays of the $B_c^+$ meson within
the PQCD framework~\cite{Liu:2009qa,Liu:2010da,Liu:2010kq,Liu:2010epa,Xiao:2011zz,Xiao:2013lia,Liu:2017cwl},
we aim to explore the perturbative dynamics in the considered channels through
observables such as the branching ratios and their relative ratios. This will provide valuable
information to aid in understanding the nature of light scalars and further investigating their internal
structure, even provide more evidences for exploring the annihilation mechanism in the weak decays of charm and bottom quarks.
Theoretically, the iPQCD formalism is now more self-consistent at leading order by incorporating the new Sudakov factor
that accounts for the charm quark mass effects~\cite{Liu:2018kuo,Liu:2020upy}. Furthermore, utilizing the newly proposed
$B_c^+$-meson distribution amplitude, the current leading-order iPQCD predictions on the $B_c^+ \to
J/\psi\ (n)\pi$ ($n=1,2,3$)~\cite{Liu:2023kxr} and $B_c^+ \to \chi_{cJ} \pi^+$ ($J=0,1,2$)~\cite{Liu:2025ipe}
decays are generally consistent
with the available measurements or upper limits reported by the LHCb Collaboration within uncertainties.
Frankly speaking, this phenomenology demonstrates the preliminary success of the iPQCD formalism,
which encourages its application in this work.
Experimentally, increasingly precise measurements will be conducted using the fully upgraded LHCb detector,
which will provide significant support for theoretical explorations of the dynamics of light scalars,
as well as for further testing the reliability of this iPQCD formalism utilized in the $B_c^+$-meson decays.

This paper is organized as follows. In Sec.~\ref{sect:2},
we present a brief overview on the light scalars and tensors and the perturbative calculations
of pure annihilation decay $B_c^+ \to ST$ in the iPQCD formalism.
Then we give the numerical results and make phenomenological analyses in Sec.~\ref{sect:R&D}.
Finally, Sec.~\ref{sect:C&S} contains the main conclusions and a short summary.
	
%
%
\section{ FORMALISM and Perturbative calculations}
\label{sect:2}

\subsection{Light scalar and tensor mesons}
Before proceeding, it is necessary to provide a brief overview on
the current status of light scalars and tensors.
The light scalars considered in this work so far include $f_0(500)$, $K_0^*(700)$, $f_0(980)$, and $a_0(980)$,
as well as $f_0(1370)$, $K_0^*(1430)$, $a_0(1450)$, $f_0(1500)$ and $f_0(1710)$.
Similarly, the well-established light tensors include $f_2(1270)$, $a_2(1320)$, $K_2^*(1430)$, and $f_2'(1525)$,
which exhibit a more conventional quark--antiquark structure and serve as important benchmarks for spectroscopic classification~\cite{ParticleDataGroup:2024cfk}. For simplicity, we will adopt $a_0$, $K_0^*$, $f_0$
and $f_0^\prime$ to
denote the light scalar states $a_0(980, 1450)$, $K_0^*(700, 1430)$, $f_0(500, 1370)$ and $f_0(980, 1500)$,
as well as $a_2$, $K_2^*$, $f_2$ and $f_2^\prime$ to stand for the light tensor states $a_2(1320)$,
$K_2^*(1430)$, $f_2(1270)$ and $f_2^\prime(1525)$, respectively, in the following context, unless otherwise stated.

As indicated in the {\it Review of Particle Physics}, the light scalars above 1~GeV can be
regarded as the conventional $q\bar q$ mesons, while those below 1~GeV remain to be determined~\cite{ParticleDataGroup:2024cfk}.
That is, the inner structure of light scalars below 1~GeV is very complicated and remains controversial.
Indeed, at the experimental aspect, the CMS Collaboration's analysis strongly
favors a conventional $q\bar q$-interpretation of the resonance $f_0(980)$~\cite{CMS:2023rev}.
In contrast, the analysis by the ALICE Collaboration clearly concludes that the $K^*_0(700)$ state
has a compact tetraquark structure~\cite{ALICE:2023eyl}.
As emphasized in~\cite{Cheng:2005nb}, factorization frameworks are effective for $q\bar q$ mesons
but not for tetraquarks or other multiquark states. Consequently, this study considers these light scalar
and tensor particles as conventional two-quark mesons within the quark model.

Fortunately, light tensor states are generally regarded as having the conventional $q\bar q$ structure.
Meanwhile, the flavor-neutral states $f_2(1270)$ and $f_2^\prime(1525)$ arise from
mixing between two quark-flavor states $f_{2q}$ and $f_{2s}$ in the quark-flavor basis~\cite{Li:2000zb,Cheng:2011fk},
\beq
\bigg (  \begin{array}{c}
|f_2(1270) \rangle \\
|f_2^{\prime}(1525) \rangle
\end{array}  \bigg )  &=&
\bigg ( \begin{array}{cc}
\cos{\alpha} & -\sin{\alpha}  \\
\sin{\alpha} & \cos{\alpha}
\end{array}  \bigg )
\bigg (  \begin{array}{c}
|f_{2q}\rangle \\
|f_{2s} \rangle
\end{array} \bigg )\;,
\label{eq:5}
\eeq
similar to the well-known $\eta-\eta^\prime$ mixing in the pseudoscalar sector,
with the mixing angle $\alpha$, $f_{2q} \equiv (u\bar u + d\bar d)/\sqrt{2}$ and $f_{2s} \equiv s\bar s$,
where the masses of $f_{2q}$ and $f_{2s}$ are determined by the following relations,
\beq
m_{f_{2q}}^2&=&m_{f_2(1270)}^2\cos^2{ \alpha }+m_{{f_2^\prime}(1525)}^2\sin^2{ \alpha }\;,
\quad
m_{f_{2s}}^2=m_{f_2(1270)}^2\sin^2{ \alpha }+m_{{f_2^\prime}(1525)}^2\cos^2{ \alpha }\;.
\label{eq:6}
\eeq
While,
the $f_2(1270)$ and $f_2^\prime(1525)$ could be considered approximately as the productions
of ideal mixing, due to the fact that the mixing angle $\alpha$ lied in the range of $6^\circ$
to $10^\circ$ as implied by the related studies~\cite{Li:2000zb,Cheng:2011fk}.
That is, such small mixing leads to $f_2(1270) \simeq f_{2q}$ and $f_2^\prime(1525) \simeq f_{2s}$,
which will be adopted in this work.

As noted above, the situation for light scalars is extraordinarily complicated.
Nevertheless, by employing the assignment $q\bar q$ for light scalars, it is proposed
based on QCD sum rule calculations that
the aforementioned light scalar mesons can be classified into the following two distinct
scenarios~\cite{Cheng:2005nb}.
\begin{itemize}
\item
In scenario $S1$, all scalar mesons are interpreted as conventional two-quark states.
Specifically, the nonet below 1~GeV, including $a_0(980)$, $K_0^*(700)$, $f_0(500)$, and
$f_0(980)$, is considered to be the ground state, while the nonet around 1.5~GeV,
including $a_0(1450)$, $K_0^*(1430)$, $f_0(1370)$, and $f_0(1500)$, is viewed correspondingly
as the first orbitally excited state.

\item
In scenario $S2$, the scalar mesons in the vicinity of 1.5~GeV can be interpreted as the
lowest-lying $q\bar{q}$ states, while the lighter scalars below 1~GeV
are considered as exotic tetraquark (diquark-antidiquark) states.
\end{itemize}

More comments on the isoscalars $f_0(500, 980, 1370, 1500)$ are in order.
\begin{itemize}
\item
{\it Scalar glueball}

A prominent topic within the scalar sector is undoubtedly the remarkably
intriguing scalar glueball. But, different from the very recently reported pseudoscalar glueball~\footnote{
It is of great interest to note that, very recently, the BESIII Collaboration
declared the observation of $0^{-+}$ glueball-dominated state, that is, $X(2370)$.
The interested readers could refer to~\cite{BESIII:2026mvn} for detail.},
the scalar glueball is more challenging and to be determined in the future.
The idea of primary scalar glueball was initially proposed. However,
a controversial issue remains regarding which of the two possible candidates,
$f_0(1500)$ and $f_0(1710)$, is the true one. Two contrasting viewpoints suggested by Close {\it et al.}~\cite{Close:2005vf}
and Cheng {\it et al.}~\cite{Cheng:2015iaa} have been available
for a long time (for a review, see, e.g., Ref.~\cite{Ren:2023ebq} for detail, and references therein).
In 2021, Klempt {\it et al.} proposed an extremely different viewpoint, namely, fragmented scalar glueball~\cite{Klempt:2021nuf}. They believed that the fragmented scalar glueball contributes into
isoscalars with different probabilities. Moreover, it is found that the isoscalar $f_0$ and
$f_0^\prime$ states
considered in this work contain the fragmented scalar glueball with only few percent. Therefore,
it means that the isoscalars $f_0(500)$ and $f_0(980)$, as well as $f_0(1370)$ and $f_0(1500)$,
might mainly be an admixture of light scalar quarkonia $(u\bar u + d\bar d)/\sqrt{2}$ and
$s\bar s$ with definite fractions fitted from the increasingly precise data.
Maybe these two kinds of viewpoints, namely, primary scalar glueball and fragmented scalar glueball,
could be crosschecked by the more and more precise measurements in the future experiments.

\item
{\it $f_0(500)-f_0(980)$ mixing}

As inferred from~\cite{ParticleDataGroup:2024cfk}, experimental evidence has demonstrated
that $f_0(980)$ is not a pure $s\bar{s}$ state. Then
the scalar mesons $f_0(500)$ and $f_0(980)$ are supposed to have a mixing similar to that of
the pseudoscalar $\eta$ and $\eta^\prime$ mesons.
Therefore, the mixing of $f_0(500)-f_0(980)$ can be read as
\beq
\bigg (  \begin{array}{c}
|f_0(500) \rangle \\
|f_0(980) \rangle
\end{array}  \bigg  )  =
\bigg ( \begin{array}{cc}
\cos{\theta} & -\sin{\theta}  \\
\sin{\theta} & \cos{\theta}
\end{array}  \bigg )
\bigg (  \begin{array}{c}
|f_{0}^n\rangle \\
|f_{0}^s \rangle
\end{array} \bigg )\;,
\label{eq2}
\eeq
with the quark-flavor states $f_{0}^n = (\bar{u} u + \bar{d} d) /\sqrt{2}$, $f_{0}^s = \bar{s}s$,
and the mixing angle $\theta$~\cite{Cheng:2005nb,Cheng:2007st}.
For the mixing angle $\theta$ between $f_0(500)$ and $f_0(980)$, numerous phenomenological studies on its magnitude are available in the literature, based on experimental measurements of their decays and productions
(for a review, see e.g., Ref.~\cite{Klempt:2021nuf} for details). The
allowed range has been suggested as $25^\circ < \theta < 40^\circ$ or $140^\circ < \theta < 165^\circ$~\cite{Cheng:2005nb}, and $|\theta| < 29^\circ$ at 90\% C.L.~\cite{Stone:2013eaa}.
Moreover, a data-driven analysis of $B_{(s)}^0 \to J/\psi\ f_0(500, 980)[\to\pi^+\pi^-]$ in the PQCD
framework yields general agreement between theoretical predictions and experimental measurements when
the mixing angle is fixed at $|\theta| = 25^\circ$~\cite{Liu:2019ymi}. We will adopt
this value to calculate the related branching ratios in this work.

Additionally, the mixing between $f_0(980)$ and $a_0(980)$,
which involves a transition between isoscalar and isovector states,
has been experimentally verified~\cite{BESIII:2018ozj}.
Although a thorough understanding of $a_0(980)-f_0(980)$ mixing could help
to study the internal structure of light scalar mesons~\cite{Li:2022pkl,Wang:2016wpc,Roca:2012cv,Liu:2010id,BESIII:2010dhc,Wu:2008hx},
a long-standing problem in hadron physics,
it is not the main focus of the present work and must be left for future investigations.

\item
{\it $f_0(1370)-f_0(1500)$ mixing}

As suggested in~\cite{Klempt:2021nuf}, this mixing can be represented by a unitary transformation with the mixing angle $\varphi$ as
\beq
\bigg (  \begin{array}{c}
|f_0(1370) \rangle \\
|f_0(1500) \rangle
\end{array}  \bigg  )  &=&
\bigg ( \begin{array}{cc}
\cos{\varphi} & -\sin{\varphi}  \\
\sin{\varphi} & \cos{\varphi}
\end{array}  \bigg )
\bigg (  \begin{array}{c}
|f_{0}^N\rangle \\
|f_{0}^S \rangle
\end{array} \bigg )\;,
\label{eq:3}
\eeq
Here, $f_{0}^N$ and $f_0^S$ likewise denote the quark-flavor components $(\bar{u}u+\bar{d}d)/\sqrt{2}$ and $\bar{s}s$, respectively, 
in the scalar $f_0(1370)$ and $f_0(1500)$ mesons. And the value of $\varphi$ is adopted as $|\varphi| = (146 \pm 8)^\circ$~\cite{Klempt:2021nuf,Klempt:2021wpg,Li:2025ddc}.
\end{itemize}

\subsection{Perturbative Calculations}

We work in the $B_c^+$-meson rest frame and use the light-cone coordinates to describe the kinematic variables,
assuming that one meson moves along the plus $z$-direction and another moves along the minus $z$-direction.
Then we obtain the momenta $P_1$ and $P_{2,3}$ of the initial- and final state mesons as
\beq
P_{1}&=&\frac{m_{B_c}}{\sqrt{2}}(1,1,{\bf 0}_T)\;,
\qquad
P_{2}=\frac{m_{B_c}}{\sqrt{2}}(1-r_3^2,r_2^2,{\bf 0}_T)\;,
\qquad
P_{3}=\frac{m_{B_c}}{\sqrt{2}}(1-r_2^2,r_3^2,{\bf 0}_T)\;,
\label{eq:9}
\eeq
where $m_{B_c}$ is the $B_c^+$-meson mass and $r_{2,3}$
satisfies $r_{2,3}=m_{2,3}/m_{B_c}$ with $m_{2,3}$
being the masses of final states in the considered decays.
Here, the term involving $r_2^2\cdot r_3^2$ is indeed power-suppressed and thus
has been safely neglected.
The momenta of valence quarks in the involved hadrons are chosen as
\beq
k_1 = (x_1 P_1^+, x_1 P_1^-, \boldsymbol{k}_{1T})\;,
\qquad
k_2 = (x_2 P_2^+, x_2 P_2^-, \boldsymbol{k}_{2T})\;,
\qquad
k_3 = (x_3 P_3^+, x_3 P_3^-, \boldsymbol{k}_{3T})\;,
\label{eq:10}
\eeq
where $x_i(i=1,2,3)$ is the corresponding momentum fraction.
Due to the angular-momentum conservation, only longitudinal polarization amplitudes contribute to the considered processes.
Then, the longitudinal polarization vector of the tensor meson with momentum $P_2$ or $P_3$ could be newly constructed on
the basis of longitudinal polarization vector of the vector meson and given as follows~\cite{Liu:2017cwl},
\beq
\epsilon_2^L&=& \sqrt{\frac{2}{3}}\frac{1}{\sqrt{2(1-r_{3}^{2})} r_{2}}(1-r_{3}^{2},-r_{2}^{2}, {\bf 0}_{T})\;,
\qquad
\epsilon_3^L= \sqrt{\frac{2}{3}}\frac{1}{\sqrt{2(1-r_{2}^{2})} r_{3}}(-r_{3}^{2},1-r_{2}^{2}, {\bf 0}_{T})\;,
\label{eq:11}
\eeq
And then, the integration over $k_1^-$, $k_2^-$ and $k_3^+$ can conceptually lead to the decay amplitude in the
PQCD factorization framework~\cite{Keum:2000wi,Lu:2000em}, for example,
\beq
A(B_c^+ \to ST) &\sim &\int\!\! d x_1 d
x_2 d x_3 b_1 d b_1 b_2 d b_2 b_3 d b_3
\non && \cdot {\mathrm{Tr}}
\left [ C(t) \Phi_{B_c}(x_1, b_1) \Phi_{S}(x_2, b_2)
\Phi_{T}(x_3, b_3) H(x_i, b_i, t) S_t(x_i)e^{-S(t)} \right ]\;,
\label{eq:12}
\eeq
where $\Phi_{B_c}$, $\Phi_{S}$ and $\Phi_{T}$ represent the wave functions of $B_c$, scalar, and tensor mesons, respectively;
$x_i(i=1,2,3)$ is the momentum fraction of valence quark in the initial and final mesons,
$\boldsymbol{b}_i(i=1,2,3)$ is the conjugate-space coordinate of transverse momentum $\boldsymbol{k}_{iT}(i=1,2,3)$;
$H(x_i, b_i, t)$ is the hard kernel with $t$ being the largest running energy scale, and
$C(t)$ denotes the related Wilson coefficients.
Various double logarithms are resummed by the $\boldsymbol{k}_T$ resummation and threshold resummation
techniques, then the Sudakov factor $e^{-S(t)}$~\cite{Liu:2018kuo, Liu:2020upy,Li:2003yj,Botts:1989kf,Li:1992nu}
and the jet function $S_t(x_i)$~\cite{Li:2001ay, Li:2002mi} are naturally
obtained, which strongly suppress the long-distance (large b) contributions.

As proposed in Ref.~\cite{Liu:2018kuo}, the wave function of the $B_c^+$ meson adopted
in the iPQCD formalism can be written as
\beq
\Phi_{B_c}(x, \boldsymbol{k}_T)  &=&
\frac{i}{\sqrt{2N_c}} \biggl \{(\psl_{B_c} +m_{B_c})\gamma_5 \phi_{B_c}(x, \boldsymbol{k}_T) \biggr \}
\;,
\label{eq:8}
\eeq
where
$N_c$ is the color number, $P_{B_c}$ and $m_{B_c}$ are the momentum and mass of the $B_c^+$ meson, respectively,
and $\phi_{B_c}(x, \boldsymbol{k}_T)$ is the leading-twist distribution amplitude with
$x$ and $\boldsymbol{k}_T$ being the momentum fraction and the intrinsic transverse momentum of 
charm quark in the $B_c^+$ meson.

For the light scalar meson, the light-cone wave function has been calculated in the QCD sum rule up
to twist-3~\cite{Cheng:2005nb,Cheng:2007st,Cheng:2013fba,Li:2008tk},
\beq
\Phi_{\!S}(x) &=&
\frac{i}{ \sqrt{2 N_c}} \bigg\{\psl \phi_{\!S}(x) + m_{\!S} \phi_{\!S}^S(x) + m_{\!S} (\nsl\vsl - 1) \phi_{\!S}^T(x) \bigg\}
\;,
\label{eq:9}
\eeq
where $\phi_{S}(x)$ is the twist-2 distribution amplitude,
$\phi_{\!S}^S(x)$ and $\phi_{\!S}^T(x)$ are twist-3 ones with
$x$ the quark momentum fraction of the light scalar meson, $P$ and $m_S$
are the momentum and mass of the scalar meson, respectively, and
$n=(1,0,{\bf 0}_T)$ and $v=(0,1,{\bf 0}_T)$ are the dimensionless light-like unit vectors.

For the light tensor meson, Cheng {\it et al.} ever used the QCD sum rule to derive the two-parton light-cone
wave function up to twist-3~\cite{Cheng:2010hn, Wang:2010ni},
\beq
\Phi^{L}_{\!T}(x) &=&  \frac{1}{\sqrt{2 N_c}}  \bigg\{ m_{\!T}\, {\epsl}_{\!T}^L \phi_{\!T}(x)
+ {\epsl}_{\!T}^L \, \psl \phi_{\!T}^t(x)  + m_{\!T} \phi_{\!T}^s(x) \bigg \}
\;,
\label{eq:4}
\eeq
where $\phi_{T}(x)$ is the leading-twist distribution amplitude,
$\phi_{\!T}^t(x) $ and $\phi_{\!T}^s(x)$ are the twist-3 ones,
$m_T$ and $P$ are the momentum and mass of the tensor meson,
and $\epsilon_{\!T}^L$ denotes the longitudinal polarization vector of the tensor meson~\cite{Datta:2007yk,Liu:2023kxr}.
More details about the above-mentioned light-cone distribution amplitudes (LCDAs) can be found in the Appendix~\ref{a1}.

The effective Hamiltonian for the decay $B_c^+ \to ST$ in the standard model is given as~\cite{Buchalla:1995vs},
\beq
H_{\rm eff} &=&
{G_{\!F}\over\sqrt{2}} \bigg\{ V^*_{cb}V_{uq} [ C_1(\mu) O_1(\mu) +C_2(\mu) O_2(\mu) ] \bigg\}+ {\rm H.c.}\;,
\label{eq:13}
\eeq
in which the Fermi constant $G_{\!F}=1.16639\times 10^{-5}\ {\rm GeV}^{-2}$, $ V^*_{cb}$ and $V_{uq}$ are
the Cabibbo-Kobayashi-Maskawa (CKM) matrix elements with ``$q$" describing the light down-type quark $d$
or $s$, and $C_i(\mu)$ denotes Wilson coefficients at the renormalization scale $\mu$.
For the Wilson coefficients $C_{1,2}(\mu)$, we will adopt the leading order expressions, although the next-to-leading
order calculations already exist in the literature~\cite{Buchalla:1995vs}. This is a consistent
way to cancel the explicit dependence on $\mu$ in the theoretical formulas. For the evolution of the Wilson coefficients from the higher scale to the lower scale, we use the formulas given in~\cite{Lu:2000em} directly.
The local four-quark operators $ O_1(\mu)$ and
$O_2(\mu)$ are given as follows
\beq
O_1 & =&
\bar{q}_\alpha \gamma_\mu (1 - \gamma_5) u_\beta\; \bar{c}_\beta \gamma_\mu (1 - \gamma_5) b_\alpha \;,
\qquad
O_2  =
\bar{q}_\alpha \gamma_\mu (1 - \gamma_5) u_\alpha\; \bar{c}_\beta \gamma_\mu (1 - \gamma_5) b_\beta \;,
\label{eq:14}
\eeq
where $q=s$ for the CKM-suppressed processes with $\Delta S = 1$ and $q=d$ for the CKM-favored processes with
$\Delta S = 0$ (Here, the ``$S$" stands for the quantum strange number).

\begin{figure}[!!htb]
\centering
\includegraphics[width=0.95\textwidth]{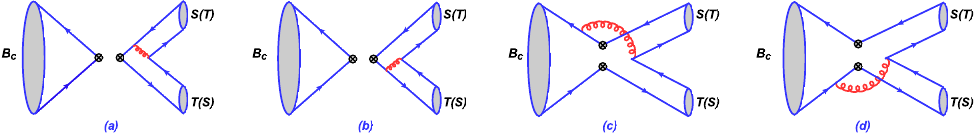}
\caption{Leading-order Feynman diagrams of the pure annihilation decay $B_c^+ \to S T$ in the iPQCD formalism, with (a) and (b) [(c) and (d)] contributing to the (non-) factorizable decay amplitudes.}
\label{f:1}
\end{figure}

Fig.~\ref{f:1} shows the leading-order Feynman diagrams for the pure annihilation
decay $B_c^+ \to ST$ in the iPQCD formalism.
Based on these diagrams, we could derive the corresponding Feynman amplitudes.
Specifically, Fig.~\ref{f:1}(a) and~\ref{f:1}(b) contribute to the factorizable
annihilation amplitudes,
while Fig.~\ref{f:1}(c) and~\ref{f:1}(d) contribute to the non-factorizable annihilation
amplitudes. We adopt $F$ and $M$ to denote the factorization formulas for the factorizable
annihilation and non-factorizable annihilation contributions, respectively.

By combining various contributions, the decay amplitude of the considered modes is hence
given in the following form,
\beq
 \mathcal{A} (B_c^+ \to S T) = V_{cb}^{\ast}V_{uq}\left\{ \Biggl(\frac{C_1}{3} + C_2\Biggr)\ F +C_1\ M \right\}  \;,
 \label{eq:DecAmp}
\eeq
where the expressions of $F$ and $M$, and the decay amplitude ${\cal A}$ for each channel
have been collected in the Appendix~\ref{a3} explicitly.

With the decay amplitude ${\cal A}$ shown in Eq.~(\ref{eq:DecAmp}) for the decay $B_c^+ \to S T$,
the branching ratio can then be expressed as follows,
\beq
{\rm BR}(B_c^+ \to S T)
&\equiv&
\tau_{B_c} \cdot \Gamma(B_c^+ \to S T) \non
&=& \tau_{B_c} \cdot \frac{G_F^2 m_{B_c}^3}{32\pi} \cdot \Phi(r_S, r_T) \cdot \left| {\cal A}(B_c^+ \to S T) \right|^2\;,
\label{eq:15}
\eeq
where $\tau_{B_c} = 0.507$ ps is the $B_c^+$-meson lifetime, $\Gamma$ is the decay width,
$r_{S(T)} = m_{S(T)}/m_{B_c}$ with $m_{S(T)}$ being the mass of light scalar (tensor) meson,
and $\Phi$ is the phase space factor of $B_c^+ \to S T$ in the form of
$\Phi(x, y) \equiv \left\{ [1 - (x + y)^2] \cdot [1 - (x - y)^2] \right\}^{1/2}$.

%
%
\section{Numerical results and discussions}
\label{sect:R&D}
	
In this section, we will present the numerical results of branching ratios and relative ratios,
associated with phenomenological analyses for those considered pure annihilation
type $B_c^+ \to ST$ decay.

The relevant QCD scale (GeV)
and mesonic masses (GeV) are collected as follows~\cite{ParticleDataGroup:2024cfk},
\beq
 \Lambda_{\overline{\mathrm{MS}}}^{(f=4)} &=& 0.250\;, \quad m_W = 80.4\;, \quad
 m_{B_c} = 6.276\;, \quad m_b = 4.8\;,
 \quad m_c = 1.5\;, \non
m_{f_0(500)} &=& 0.500\;, \quad m_{K_0^*(700)} = 0.845\;, \quad m_{f_0(980)} = 0.990\;, \quad m_{a_0(980)} = 0.980\;, \non
m_{f_2(1270)} &=& 1.275\;, \quad m_{a_2(1320)} = 1.318\;, \quad m_{f_0(1370)} = 1.350\;, \quad m_{K_0^*(1430)} = 1.425\;, \non
m_{K_2^*(1430)} &=& 1.427\;, \quad m_{a_0(1450)} = 1.439\;, \quad m_{f_0(1500)} = 1.522\;, \quad m_{f_2^{\prime}(1525)} = 1.517\;,
\label{eq:mass}
\eeq
and the CKM matrix elements adopted in this work are as the following \cite{ParticleDataGroup:2024cfk},
\beq
|V_{cb}| &=& 0.041\;, \qquad    |V_{ud}|=0.974\;, \qquad  |V_{us}|= 0.224\;.
\label{eq:ckm}
\eeq

Due to the large factor induced by the CKM matrix elements, that is, $|V_{ud}/V_{us}|^2 \approx 19$,
it is evident that the branching ratios for the $\Delta S = 0 $ processes are generally larger than
those for the $ \Delta S = 1 $ ones. In particular, for some CKM-favored processes,
the branching ratios can reach $ \mathcal{O}(10^{-6}) $ within uncertainties.
As argued in~\cite{Descotes-Genon:2009eui}, these values are expected to be experimentally
verified at the LHCb detector. Of course, it should be pointed out that the less constrained
parameters such as decay constants and Gegenbauer moments of the considered tensors and scalars
would lead to large uncertainties theoretically. We are looking forward to the precise
inputs about these mentioned hadronic parameters obtained in theory and/or
experiment to promote the precision of theoretical predictions in the PQCD factorization framework.

\subsection{\boldmath $B_c^+ \to a_0 T$}

The $B_c^+ \to a_0 T$ decay has ten modes, namely, the $\Delta S =0$ processes
$B_c^+ \to a_0^+ a_2^0$, $B_c^+ \to a_0^0 a_2^+$,
and $B_c^+ \to a_0^+ f_2$, while the $\Delta S =1$ ones
$B_c^+ \to a_0^+ K_2^{*0}$ and $B_c^+ \to a_0^0 K_2^{*+}$,
respectively. Since $f_2$ and $f_2^\prime$ have been regarded as
the pure tensor flavor states $f_{2q}$ and $f_{2s}$, then the $B_c^+ \to a_0^+ f_2^\prime$
mode is forbidden in the standard model naturally. The numerical results of
the $B_c^+ \to a_0 T$ branching ratios predicted
in the iPQCD formalism have been presented in Tables~\ref{t:1} and \ref{t:2} explicitly.

\begin{table}[hbt]
\caption{Branching ratios of the decay $B_c^+ \to a_0(980) T$ in the iPQCD formalism. The main uncertainties
arise from the related hadronic parameters, specifically, the shape parameter $\beta_{B_c}$, the combined
decay constant $f_i$, and the Gegenbauer moment $B_i(i = 1,3)$, respectively.}
\label{t:1}
\begin{center}\vspace{-0.3cm}{\footnotesize
\begin{tabular}{c|c }
\hline
Decay modes ($\Delta S=0$)     & Branching ratios  \\
\hline
$B_c^+ \to a_0(980)^{+} a_2(1320)^0$
& $3.55^{+0.51}_{-0.49}(\beta_{B_c})
^{+0.86}_{-0.67}(f_i)
^{+2.17}_{-1.73}(B_i)
\times 10^{-8}$ \\
\hline
$B_c^+ \to a_0(980)^0  a_2(1320)^{+}$
& $3.43^{+0.50}_{-0.48}(\beta_{B_c})
^{+0.84}_{-0.65}(f_i)
^{+2.14}_{-1.71}(B_i)
\times 10^{-8}$ \\
\hline
$B_c^+ \to a_0(980)^{+} f_2(1270)$
& $
0.76
^{+0.15}_{-0.11}(\beta_{B_c})
^{+0.13}_{-0.12}(f_i)
^{+0.15}_{-0.14}(B_i)
\times 10^{-6}$ \\
\hline
Decay modes ($\Delta S=1$)  & Branching ratios \\
\hline
$B_c^+ \to a_0(980)^+ {K_2^*}(1430)^0$
& $4.54^{+0.76}_{+0.60}(\beta_{B_c})
^{+0.66}_{-0.64}(f_i)
^{+1.04}_{-0.92}(B_i)
\times 10^{-8}$ \\
\hline
$B_c^+ \to a_0(980)^0  {K_2^*}(1430)^+$
& $2.26^{+0.38}_{-0.30}(\beta_{B_c})
^{+0.38}_{-0.37}(f_i)
^{+0.52}_{-0.45}(B_i)
\times 10^{-8}$ \\
\hline
\end{tabular}}
\end{center}
\end{table}

Based on the $B_c^+ \to a_0(980) a_2$ and $B_c^+ \to a_0(980)^+ f_2$ branching ratios (BRs)
presented in Table~\ref{t:1}, several discussions are in order.
\begin{itemize}
\item
The relation ${\rm BR}(B_c^+ \to a_0(980)^+ a_2^0) \simeq {\rm BR}(B_c^+ \to a_0(980)^0 a_2^+)$
within uncertainties is obtained. While, only in terms of the central values,
the former branching ratio is slightly larger than the latter one, which is due to
the fact that the vector decay constant $f_{a_0(980)^0} = 0$ satisfying the invariance of charge conjugation or the conservation of vector current, though $f_{a_0(980)^+}$ is tiny.

\item
The $B_c^+ \to a_0(980)^+ f_2$ branching ratio is approximately $10^{-6}$ within uncertainties. In principle,
it could be probed in the future experiments at LHC. Meanwhile, an interesting ratio between the
$B_c^+ \to a_0(980)^+ f_2$ and $B_c^+ \to a_0(980)^+ a_2^0$ branching ratios could be derived
as,
\beq
\frac{{\rm BR}(B_c^+ \to a_0(980)^+ f_2)}{{\rm BR}(B_c^+ \to a_0(980)^+ a_2^0)}
&=& 21.4^{+12.8}_{-5.6} \;,
\label{eq:r-fa2-a98-s1}
\eeq
This indicates that there are significantly constructive (destructive) interferences
between the contributions coming from $u\bar u$ and $d\bar d$ components
in the former (latter) modes. The large branching ratio in association with this relative ratio
could help to explore the involved hadron dynamics of $a_0(980)$.

\item
Note that, the $B_c^+ \to a_0(980) a_2$ and $B_c^+ \to a_0(980)^+ K_2^{*0}$ branching
ratios listed in Table~\ref{t:1} are consistent with each other within large uncertainties,
however, the former (latter) is CKM-favored (suppressed). The reason is that
the significant cancelation is happened between the contributions induced by the $u\bar u$ and $d\bar d$
components in both $a_0(980)^0$ and $a_2(1320)^0$ states. The future measurements might
help us to explore the interesting QCD dynamics in these decays.

\end{itemize}

\begin{table}[htb]
\caption{
Same as Table~\ref{t:1} but for the decays $B_c^+ \to a_0(1450) T$, where the upper (lower) entry
corresponds to the $a_0(1450)$ in scenario 1 (2). }
\label{t:2}
\begin{center}\vspace{-0.3cm}{\footnotesize
\begin{tabular}{c|c}
\hline
Decay modes ($\Delta S=0$) & Branching ratios \\
\hline
\multirow{2}{*}{$B_c^+ \to a_0(1450)^{+} a_2(1320)^{0}$}
& $
0.88^{+0.08}_{-0.08}(\beta_{B_c})
^{+0.31}_{-0.27}(f_i)
^{+0.19}_{-0.14}(B_i)
\times 10^{-6}$ \\
& $1.76^{+0.04}_{-0.03}(\beta_{B_c})
^{+0.72}_{-0.63}(f_i)
^{+0.21}_{-0.20}(B_i)
\times 10^{-6}$ \\
\hline
\multirow{2}{*}{$B_c^+ \to a_0(1450)^0 a_2(1320)^+$}
& $
0.88^{+0.08}_{-0.08}(\beta_{B_c})
^{+0.37}_{-0.32}(f_i)
^{+0.19}_{-0.14}(B_i)
\times 10^{-6}$ \\
& $1.76^{+0.03}_{-0.03}(\beta_{B_c})
^{+0.72}_{-0.63}(f_i)
^{+0.21}_{-0.21}(B_i)
\times 10^{-6}$ \\
\hline
\multirow{2}{*}{$B_c^+ \to a_0(1450)^{+} f_2(1270)$}
& $0.45^{+0.04}_{-0.04}(\beta_{B_c})
^{+0.12}_{-0.11}(f_i)
^{+0.13}_{-0.09}(B_i)
\times 10^{-6}$ \\
& $
0.86^{+0.09}_{-0.09}(\beta_{B_c})
^{+0.23}_{-0.21}(f_i)
^{+0.27}_{-0.22}(B_i)
\times 10^{-6}$ \\
\hline
Decay modes ($\Delta S=1$)  & Branching ratios \\
\hline
\multirow{2}{*}{$B_c^+ \to a_0(1450)^+ {K^*_2}(1430)^0$}
& $
0.99^{+0.10}_{-0.09}(\beta_{B_c})
^{+0.24}_{-0.22}(f_i)
^{+0.19}_{-0.20}(B_i)
\times 10^{-7}$ \\
& $1.29^{+0.16}_{-0.11}(\beta_{B_c})
^{+0.42}_{-0.38}(f_i)
^{+1.00}_{-0.67}(B_i)
\times 10^{-8}$ \\
\hline
\multirow{2}{*}{$B_c^+ \to a_0(1450)^0 {K^*_2}(1430)^+$}
& $0.49^{+0.00}_{-0.05}(\beta_{B_c})
^{+0.12}_{-0.11}(f_i)
^{+0.09}_{-0.10}(B_i)
\times 10^{-7}$ \\
& $
0.65^{+0.06}_{-0.08}(\beta_{B_c})
^{+0.16}_{-0.14}(f_i)
^{+0.49}_{-0.33}(B_i)
\times 10^{-8}$ \\
\hline
\end{tabular}}
\end{center}
\end{table}

While, as seen in Table~\ref{t:2}, the numerical results of the $B_c^+ \to a_0(1450) T$
branching ratios present a bit different phenomena.
\begin{itemize}
\item
Contrary to the $B_c^+ \to a_0(980) a_2$ and
$B_c^+ \to a_0(980)^+ f_2$ decays, the constructive (destructive) interferences occurring
in the $B_c^+ \to a_0(1450) a_2$
$( B_c^+ \to a_0(1450)^+ f_2 )$ ones between the $u\bar u$
and $d\bar d$ components of $a_0(1450)\ ( f_2 )$ lead to the evidently larger (smaller)
branching ratios in each scenario.
Meanwhile,
the similar ratios between the $B_c^+ \to a_0(1450)^+ a_2^0$
and $B_c^+ \to a_0(1450)^+ f_2$ branching ratios in both $S1$ and $S2$ could be deduced,
\beq
\frac{{\rm BR}(B_c^+ \to a_0(1450)^+ a_2^0)
}{{\rm BR}(B_c^+ \to a_0(1450)^+ f_2)}{\bigg |}_{S1}
&=&
1.95^{+0.19}_{-0.21} \;,
\quad
\frac{{\rm BR}(B_c^+ \to a_0(1450)^+ a_2^0)
}{{\rm BR}(B_c^+ \to a_0(1450)^+ f_2)}{\bigg |}_{S2}
=
2.05^{+0.49}_{-0.47} \;,
\label{eq:r-af2-a145}
\eeq
which can be tested in the future LHC experiments.

\item
The branching ratios of $B_c^+ \to a_0(1450)^+ a_2^0$ and $B_c^+ \to a_0(1450)^0 a_2^+$ are
equal to each other with the value $(0.88^{+0.37}_{-0.31})\times 10^{-6}$
within uncertainties and could be tested in the LHC experiments. Note that, $a_0(1450)$ is
the first excited state of $a_0(980)$ in $S1$. Hence, the future measurements on the ratio
between the $B_c^+ \to a_0(1450) a_2$
and $B_c^+ \to a_0(980) a_2$ branching ratios in $S1$,
\beq
\frac{{\rm BR}(B_c^+ \to a_0(1450)^+ a_2^0)}{{\rm BR}(B_c^+ \to a_0(980)^+ a_2^0)}{\bigg |}_{S1}
&=& 24.8^{+16.1}_{-7.2}  \;,
\quad
\frac{{\rm BR}(B_c^+ \to a_0(1450)^0 a_2^+)}{{\rm BR}(B_c^+ \to a_0(980)^0 a_2^+)}{\bigg |}_{S1}
= 25.7^{+17.7}_{-8.7} \;,
\eeq
and on that between the $B_c^+ \to a_0(1450)^+ f_2$
and $B_c^+ \to a_0(980)^+ f_2$ branching ratios also in $S1$,
\beq
\frac{{\rm BR}(B_c^+ \to a_0(1450)^+ f_2)}{{\rm BR}(B_c^+ \to a_0(980)^+ f_2)}{\bigg |}_{S1}
&=& 0.59^{+0.08}_{-0.07}  \;.
\eeq
may help us to understand the QCD dynamics of $a_0(980)$ and $a_0(1450)$.

\item
For $a_0(1450)$ as the lowest-lying scalar meson in $S2$,
the $B_c^+ \to a_0(1450) a_2$ decays
have large branching ratios in the order of $10^{-6}$.
Meanwhile, even if suffered from destructive interferences, the $B_c^+ \to a_0(1450)^+ f_2$ branching ratio within
uncertainties is also measurable at the LHCb detector. These numerical results obtained in the iPQCD formalism
could be tested in the future experiments.

\item
The $B_c^+ \to a_0(1450) a_2$ and $B_c^+ \to a_0(1450)^+ f_2$ branching ratios in both scenarios
$S1$ and $S2$ are consistent with each other, respectively, within
still large theoretical errors. It may be challenging to identify the preferred scenario based
solely on measurements of these individual branching ratios. However, by incorporating the ratios presented in
Eqs.~(\ref{eq:r-fa2-a98-s1}) and~(\ref{eq:r-af2-a145}), which exhibit
highly different values, future measurements of these iPQCD predictions for the
$B_c^+ \to a_0 (a_2, f_2)$ processes could be used to indicate the favored scenario and
to further explore the QCD dynamics in these decays as well as in the light mesons.

\item
It is also noteworthy that, in light of the central values for the $B_c^+ \to a_0(1450) T$
branching ratios as shown in Table~\ref{t:2}, the results in $S1$ are obviously smaller
(larger) than those
in $S2$ for the $\Delta S =0 (1)$ processes. This indicates markedly different dynamics in
the related decay channels, which will be addressed through future experimental explorations.

\end{itemize}

Moreover, as shown in Tables~\ref{t:1} and~\ref{t:2}, it is clear that the good isospin symmetry
could be observed in the $\Delta S=1$ decay $B_c^+ \to a_0
K_2^*$ through ${\rm BR}(B_c^+ \to a_0^0 K_2^{*+}) \simeq 2 \cdot
{\rm BR}(B_c^+ \to a_0^+ K_2^{*0})$, apart from the negligible differences
between the vector decay constants $f_{a_0^0}$ and $f_{a_0^+}$.

\subsection{\boldmath $B_c^+ \to K_0^* T$}

Next, we turn to analyze the decay $B_c^+ \to K_0^* T$, in which
twelve channels are contained, specifically, the $\Delta S =0$ processes
$B_c^+ \to \bar{K}_0^{*0} K_2^{*+}$ and
$B_c^+ \to \bar{K}_2^{*0} K_0^{*+}$,  and the $\Delta S =1$
ones $B_c^+ \to K_0^{*+} (a_2^0, f_2, f_2^\prime)$
and $B_c^+ \to a_2^+ K_0^{*0}$, respectively.
Their branching ratios predicted in the iPQCD formalism can be found
in Tables~\ref{t:3} and \ref{t:4} clearly.

\begin{table}[htb]
\caption{
Same as Table~\ref{t:1} but for the decay $B_c^+ \to K_0^*(700)T$. }
\label{t:3}
\begin{center}\vspace{-0.3cm}{\footnotesize
\begin{tabular}{c|c }
\hline
Decay modes($\Delta S=0$)     & Branching ratios   \\
\hline
$B_c^+ \to {\bar{K}_0^*}(700)^0 {K_2^*}(1430)^+$
& $1.52^{+0.35}_{-0.25}(\beta_{B_c})
^{+0.24}_{-0.23}(f_i)
^{+0.25}_{-0.24}(B_i)
\times 10^{-7}$ \\
\hline
$B_c^+ \to {K_0^*}(700)^+  {\bar{K}_2^*}(1430)^0$
& $
0.85^{+0.14}_{-0.12}(\beta_{B_c})
^{+0.13}_{-0.12}(f_i)
^{+0.22}_{-0.19}(B_i)
\times 10^{-6}$ \\
\hline
Decay modes($\Delta S=1$)  & Branching ratios  \\
\hline
$B_c^+ \to {K_0^*}(700)^0 a_2(1320)^+$
& $5.07^{+1.29}_{-1.04}(\beta_{B_c})
^{+0.97}_{-0.92}(f_i)
^{+1.26}_{-1.23}(B_i)
\times 10^{-9}$ \\
\hline
$B_c^+ \to  {K_0^*}(700)^+ a_2(1320)^0$
& $2.52^{+0.65}_{-0.51}(\beta_{B_c})
^{+0.51}_{-0.49}(f_i)
^{+0.64}_{-0.62}(B_i)
\times 10^{-9}$ \\
\hline
$B_c^+ \to {K_0^*}(700)^+ f_2(1270)$
& $6.14^{+1.40}_{-1.11}(\beta_{B_c})
^{+1.36}_{-1.27}(f_i)
^{+1.52}_{-1.45}(B_i)
\times 10^{-9}$ \\
\hline
$B_c^+ \to  {K_0^*}(700)^+ f_2^{\prime}(1525)$
& $5.54^{+0.91}_{-0.74}(\beta_{B_c})
^{+0.77}_{-0.73}(f_i)
^{+1.37}_{-1.22}(B_i)
\times 10^{-8}$ \\
\hline
\end{tabular}}
\end{center}
\end{table}

From the numerical results of branching ratios presented in Table~\ref{t:3}, some remarks are in order.
\begin{itemize}
\item
Distinctly different QCD dynamics exhibits through the branching ratios of the $B_c^+ \to {\bar{K}_0^*}(700)^0 K_2^{*+}$
and $B_c^+ \to {K_0^*}(700)^+ \bar{K}_2^{*0}$ decays, although both of which are the CKM-favored modes.
Additionally, the relative ratio between the iPQCD branching ratios of $B_c^+ \to {K_0^*}(700)^+ \bar{K}_2^{*0}$ and
$B_c^+ \to {\bar{K}_0^*}(700)^0 K_2^{*+}$ is derived as,
\beq
\frac{{\rm BR}(B_c^+ \to {K_0^*}(700)^+ \bar{K}_2^{*0})}{{\rm BR}(B_c^+ \to {\bar{K}_0^*}(700)^0 K_2^{*+})}
=5.59^{+0.49}_{-0.52}
\eeq
Combined with the prediction of large branching ratio around $10^{-6}$ for $B_c^+ \to {K_0^*}(700)^+ \bar{K}_2^{*0}$,
the future experimental tests could provide evidences for exploring the inner structure of $K_0^*(700)$.

\item
Though those branching ratios of the $B_c^+ \to K_0^*(700) (a_2, f_2 ,f_2^\prime)$ modes are too apparently 
small to be measured soon in experiments,
some interesting relations could be observed clearly. Firstly, the isospin symmetry is still obeyed well in the $B_c^+ \to K_0^*(700)
a_2$ decays. Secondly, different dynamics arising from the similar components $u\bar u$ in the tensor states $a_2^0$
and $f_2$ lead to a bit smaller (larger) $B_c^+ \to {K_0^*}(700)^+ a_2^0 (B_c^+ \to {K_0^*}(700)^+ f_2)$ branching ratio in the iPQCD formalism, which may give some hints to understand these two light tensors through the following ratio,
\beq
\frac{{\rm BR}(B_c^+ \to {K_0^*}(700)^+ f_2)}{{\rm BR}(B_c^+ \to {K_0^*}(700)^+ a_2^0)}
&=&
2.44^{+0.08}_{-0.07}
\;,
\eeq
Finally, the ratio between the $B_c^+ \to {K_0^*}(700)^+ f_2$ and $B_c^+ \to {K_0^*}(700)^+ f_2^\prime$ branching ratios,
\beq
\frac{{\rm BR}(B_c^+ \to {K_0^*}(700)^+ f_2)}{{\rm BR}(B_c^+ \to {K_0^*}(700)^+ f_2^\prime)}
&=&
0.11^{+0.01}_{-0.01} \;,
\eeq
could help measure the exact mixing between those two tensor quark-flavor states $f_{2n}$ and $f_{2s}$, in association with the
future measurements with high precision.

\end{itemize}

\begin{table}[ht]
\caption{
Same as Table~\ref{t:1} but for the decay $B_c^+ \to K_0^*(1430) T$, in which the upper (lower) entry
corresponds to the scalar meson ${K_0^*}(1430)$ in scenario 1 (2).}
\label{t:4}
\begin{center}\vspace{-0.3cm}{\footnotesize
\begin{tabular}{c|c}
\hline
Decay modes ($\Delta S=0$) & Branching ratios \\
\hline
\multirow{2}{*}{$B_c^+ \to {\bar{K}^{*}_0}(1430)^0  {K^{*}_2}(1430)^+$}
& $2.10^{+0.04}_{-0.11}(\beta_{B_c})
^{+0.79}_{-0.70}(f_i)
^{+0.38}_{-0.41}(B_i)
\times 10^{-7}$ \\
& $1.69^{+0.15}_{-0.14}(\beta_{B_c})
^{+0.51}_{-0.46}(f_i)
^{+0.32}_{-0.31}(B_i)
\times 10^{-6}$ \\
\hline
\multirow{2}{*}{$B_c^+ \to {\bar{K}^{*}_2}(1430)^0 {K^{*}_0}(1430)^+$}
& $1.52^{+0.11}_{-0.11}(\beta_{B_c})
^{+0.36}_{-0.32}(f_i)
^{+0.15}_{-0.14}(B_i)
\times 10^{-6}$ \\
& $1.27^{+0.27}_{-0.29}(\beta_{B_c})
^{+0.35}_{-0.28}(f_i)
^{+2.38}_{-1.06}(B_i)
\times 10^{-7}$ \\
\hline
Decay modes ($\Delta S=1$)  & Branching ratios \\
\hline
\multirow{2}{*}{$B_c^+ \to {K^*_0}(1430)^0  a_2(1320)^+$}
& $1.15^{+0.04}_{-0.06}(\beta_{B_c})
^{+0.55}_{-0.46}(f_i)
^{+0.19}_{-0.19}(B_i)
\times 10^{-8}$ \\
& $1.01^{+0.07}_{-0.07}(\beta_{B_c})
^{+0.36}_{-0.32}(f_i)
^{+0.18}_{-0.18}(B_i)
\times 10^{-7}$ \\
\hline
\multirow{2}{*}{$B_c^+ \to {K^{*}_0}(1430)^+  a_2(1320)^0$}
& $5.72^{+0.22}_{-0.28}(\beta_{B_c})
^{+2.76}_{-2.27}(f_i)
^{+0.96}_{-0.92}(B_i)
\times 10^{-9}$ \\
& $5.01^{+0.36}_{-0.33}(\beta_{B_c})
^{+1.50}_{-1.34}(f_i)
^{+0.92}_{-0.92}(B_i)
\times 10^{-8}$ \\
\hline
\multirow{2}{*}{$B_c^+ \to {K^{*}_0}(1430)^+ f_2(1270)$}
& $1.44^{+0.05}_{-0.06}(\beta_{B_c})
^{+0.72}_{-0.59}(f_i)
^{+0.21}_{-0.20}(B_i)
\times 10^{-8}$ \\
& $1.14^{+0.07}_{-0.07}(\beta_{B_c})
^{+0.43}_{-0.38}(f_i)
^{+0.20}_{-0.19}(B_i)
\times 10^{-7}$ \\
\hline
\multirow{2}{*}{$B_c^+ \to {K^{*}_0}(1430)^+ f^\prime_2(1525)$}
& $7.64^{+0.57}_{-0.58}(\beta_{B_c})
^{+1.71}_{-1.55}(f_i)
^{+0.79}_{-0.80}(B_i)
\times 10^{-8}$ \\
& $7.91^{+0.77}_{-1.15}(\beta_{B_c})
^{+2.18}_{-1.89}(f_i)
^{+12.6}_{-6.51}(B_i)
\times 10^{-9}$ \\
\hline
\end{tabular}}
\end{center}
\end{table}

According to the numerical results for the $B_c^+ \to K_0^*(1430) T$ branching ratios
predicted in iPQCD formalism, we can find that
\begin{itemize}
\item
The $B_c^+ \to {\bar{K}^*_0}(1430)^0  K^{*+}_2$ and $B_c^+ \to \bar{K}^{*0}_2  {K^*_0}(1430)^+$
branching ratios exhibit the very contrary patterns in scenarios
1 and 2. More specifically, like those $B_c^+ \to K_0^*(700) K_2^*$ modes, the $B_c^+ \to {\bar{K}^*_0}(1430)^0  K^{*+}_2
( B_c^+ \to \bar{K}^{*0}_2  {K^*_0}(1430)^+ )$ branching ratio is evidently small (large) around $10^{-7} (10^{-6})$ calculated in $S1$.
However, the situation is completely reversed in $S2$, namely, the $B_c^+ \to {\bar{K}^*_0}(1430)^0  K^{*+}_2
( B_c^+ \to \bar{K}^{*0}_2 {K^*_0}(1430)^+ )$ branching ratio is in the order of $10^{-6} (10^{-7})$, as clearly shown in Table~\ref{t:4}.
The large branching ratios such as $(1.69^{+0.62}_{-0.57}) \times 10^{-6}$ for $B_c^+ \to {\bar{K}^*_0}(1430)^0  K^{*+}_2$ in $S2$ and
$(1.52^{+0.40}_{-0.37}) \times 10^{-6}$ for $B_c^+ \to \bar{K}^{*0}_2  {K^*_0}(1430)^+$ in $S1$, associated with the
following two interesting ratios,
\beq
\frac{{\rm BR}(B_c^+ \to {\bar{K}^*_0}(1430)^0  K^{*+}_2)}{{\rm BR}(B_c^+ \to \bar{K}^{*0}_2  {K^*_0}(1430)^+)}
&=&
0.14^{+0.02}_{-0.03} \;,
\eeq
in $S1$ and
\beq
\frac{{\rm BR}(B_c^+ \to {\bar{K}^*_0}(1430)^0  K^{*+}_2)}{{\rm BR}(B_c^+ \to \bar{K}^{*0}_2  {K^*_0}(1430)^+)}
&=&
13.3^{+52.5}_{-8.0} \;,
\eeq
in $S2$, will be tested by the experiments performed at the newly upgraded LHCb detector. The future measurements
with good precision on these iPQCD numerical values could provide important evidences to ascertain the preferred scenario
for the nature of $K_0^*(1430)$ and the reliability of the employed iPQCD formalism for $B_c$-meson decays.

\item
More ratios are derived for exploring the correlations between the ground $K_0^*(700)$ and the first-excited
$K_0^*(1430)$ scalar mesons in $S1$,
\beq
\frac{{\rm BR}(B_c^+ \to {\bar{K}^*_0}(1430)^0  K^{*+}_2)}{{\rm BR}(B_c^+ \to {\bar{K}^*_0}(700)^0  K^{*+}_2)}
&=&
1.38^{+0.32}_{-0.38} \;,
\qquad
\frac{{\rm BR}(B_c^+ \to \bar{K}^{*0}_2  {K^*_0}(1430)^+)}{{\rm BR}(B_c^+ \to \bar{K}^{*0}_2  {K^*_0}(700)^+)}
=
1.79^{+0.36}_{-0.31} \;,
\\
\frac{{\rm BR}(B_c^+ \to {K^*_0}(1430)^0  a_2^+)}{{\rm BR}(B_c^+ \to {K^*_0}(700)^0  a_2^+)}
&=&
2.27^{+0.62}_{-0.74} \;,
\qquad
\frac{{\rm BR}(B_c^+ \to a_2^0  {K^*_0}(1430)^+)}{{\rm BR}(B_c^+ \to a_2^0  {K^*_0}(700)^+)}
=
2.27^{+0.74}_{-0.71} \;.
\eeq
These ratios await the future experimental tests.

\item
For the $\Delta S =1$ decays, unfortunately, their branching ratios have such small values that they cannot be tested very soon experimentally. However, it is of great interest to note that the branching ratios of $B_c^+ \to K_0^*(1430) a_2$ and $B_c^+ \to
{K_0^*}(1430)^+ f_2$ ($B_c^+ \to {K_0^*}(1430)^+ f_2^\prime$) in $S1$ are generally
smaller (larger) than those in $S2$ with the factor around $10$, as seen in the last four lines of Table~\ref{t:4}.
Moreover, the relative ratios such as,
\beq
\frac{{\rm BR}(B_c^+ \to {K_0^*}(1430)^+ f_2)}{{\rm BR}(B_c^+ \to {K_0^*}(1430)^+ a_2^0)}
&=&
2.52^{+0.07}_{-0.08} \;,
\qquad
\frac{{\rm BR}(B_c^+ \to {K_0^*}(1430)^+ f_2)}{{\rm BR}(B_c^+ \to {K_0^*}(1430)^+ f_2^\prime)}
=
0.19^{+0.04}_{-0.05} \;,
\eeq
in $S1$, while
\beq
\frac{{\rm BR}(B_c^+ \to {K_0^*}(1430)^+ f_2)}{{\rm BR}(B_c^+ \to {K_0^*}(1430)^+ a_2^0)}
&=&
2.28^{+0.14}_{-0.21} \;,
\qquad
\frac{{\rm BR}(B_c^+ \to {K_0^*}(1430)^+ f_2)}{{\rm BR}(B_c^+ \to {K_0^*}(1430)^+ f_2^\prime)}
=
14.4^{+53.5}_{-8.1} \;,
\eeq
in $S2$, are deduced by employing the iPQCD branching ratios for the $B_c^+ \to {K_0^*}(1430)^+ a_2^0$, $B_c^+ \to {K_0^*}(1430)^+ f_2$,
and $B_c^+ \to {K_0^*}(1430)^+ f_2^\prime$ decays, which could help further understand the dynamics between $a_2$ and
$f_2$, and find the potential deviations to the ideal mixing of $f_2$ and $f_2^\prime$.
\end{itemize}

Frankly speaking, further theoretical and experimental investigations are necessary to elucidate the underlying
mechanisms contributing to these phenomena.

\subsection{\boldmath $B_c^+ \to (f_0, f_0^\prime) T$}
Finally, we constrain ourselves to the discussions of decays $B_c^+ \to f_0 T$ and $f_0^\prime T$,
which contain four $\Delta S =0$ processes $B_c^+ \to (f_0, f_0^\prime)
a_2^+$ and four $\Delta S=1$ ones $B_c^+ \to  (f_0, f_0^\prime)
K_2^{*+}$, respectively. The predictions of their branching ratios in the iPQCD formalism
are shown in Tables~\ref{t:5} and \ref{t:6}.

\begin{table}[hbt]
\caption{
Same as Table~\ref{t:1} but for the decays $B_c^+ \to f_0(500, 980)T$.}
\label{t:5}
\begin{center}\vspace{-0.3cm}{\footnotesize
\begin{tabular}{c|c}
\hline
Decay modes ($\Delta S=0 $)     & Branching ratios  \\
\hline
$B_c^+ \to f_0(500) a_2(1320)^+$
& $1.71^{+0.13}_{-0.14}(\beta_{B_c})
^{+0.41}_{-0.38}(f_i)
^{+0.27}_{-0.19}(B_i)
\times 10^{-6}$  \\
\hline
$B_c^+ \to f_0(980)  a_2(1320)^+$
& $3.64^{+0.28}_{-0.30}(\beta_{B_c})
^{+0.87}_{-0.80}(f_i)
^{+0.58}_{-0.40}(B_i)
\times 10^{-7}$ \\
 \hline
Decay modes ($\Delta S=1 $)  & Branching ratios \\
\hline
$B_c^+ \to f_0(500) {K_2^*}(1430)^+$
& $4.57^{+0.00}_{-0.12}(\beta_{B_c})
^{+1.10}_{-1.00}(f_i)
^{+0.54}_{-0.33}(B_i)
\times 10^{-8}$ \\
\hline
$B_c^+ \to f_0(980)  {K_2^*}(1430)^+$
& $3.63^{+0.43}_{-0.39}(\beta_{B_c})
^{+0.65}_{-0.60}(f_i)
^{+0.57}_{-0.39}(B_i)
\times 10^{-8}$ \\
\hline
\end{tabular}}
\end{center}
\end{table}

Generally, the CKM-suppressed $B_c^+ \to (f_0, f_0^\prime) K_2^{*+}$ modes have small branching ratios
around $10^{-8} - 10^{-7}$ within uncertainties as anticipated. For the CKM-favored
$B_c^+ \to (f_0, f_0^\prime) a_2^+$ decays,
it is noted that the $B_c^+ \to f_0 a_2^+$ modes have large branching ratios in the order of $10^{-6}$, however,
the $B_c^+ \to f_0^\prime a_2^+$ modes have slightly small branching ratios below $10^{-6}$ within uncertainties.
The underlying reason is that the $f_0$-mesons have a bit rich quark-flavor
$u\bar u$ and $d\bar d$ components in the current mixing patterns. 
In other words, these decay modes could provide constraints to the mixing
angle $\theta$ and $\varphi$. These observations can be confronted with future tests in the relevant experiments.

For the $B_c^+ \to f_0(500, 980) a_2^+$ decays, the relative ratio derived from those two branching ratios
shown in Table~\ref{t:5},
\beq
\frac{{\rm BR}(B_c^+ \to f_0(500) a_2^+)}{{\rm BR}(B_c^+ \to f_0(980) a_2^+)} &=&
4.70^{+0.00}_{-0.02}\;,
\eeq
in association with the large branching ratio, namely, ${\rm BR}(B_c^+ \to f_0(980) a_2^+) = (1.71^{+0.51}_{-0.44})\times 10^{-6}$,
could help constrain the mixing angle $\theta$ between the mixing of $f_0(500)$ and $f_0(980)$
through the following relation,
\beq
\frac{\Phi(r_{f_0(980)}, r_{a_2})}{\Phi(r_{f_0(500)}, r_{a_2})} \cdot
\frac{{\rm BR}(B_c^+ \to f_0(500) a_2^+)}{{\rm BR}(B_c^+ \to f_0(980) a_2^+)}
&=& \cot^2\!\theta\;,
\eeq
Certainly, the information of another mixing angle $\varphi$ between $f_0(1370)$ and $f_0(1500)$ could
also be extracted from
\beq
\frac{\Phi(r_{f_0(1500)}, r_{a_2})}{\Phi(r_{f_0(1370)}, r_{a_2})} \cdot
\frac{{\rm BR}(B_c^+ \to f_0(1370) a_2^+)}{{\rm BR}(B_c^+ \to f_0(1500) a_2^+)}
&=& \cot^2\!\varphi\;,
\eeq
in a similar way with the ${\cal O}(10^{-6})$ branching ratios. As shown in Table~\ref{t:6},
it is evident that the value of $\varphi$ deduced from the branching ratios is scenario
independent, as it should be.

\begin{table}[htb]
\caption{
Same as Table~\ref{t:1} but for the decays $B_c^+ \to f_0(1370, 1500)T$, in which the upper (lower) entry
corresponds to the scalars $f_0(1370, 1500)$ in scenario 1 (2).}
\label{t:6}
\begin{center}\vspace{-0.3cm}{\footnotesize
\begin{tabular}{c|c}
\hline
Decay modes ($\Delta S=0$)     &  Branching ratios \\
\hline
\multirow{2}{*}{$B_c^+ \to f_0(1370) a_2(1320)^+$}
& $0.93^{+0.10}_{-0.09}(\beta_{B_c})
^{+0.24}_{-0.22}(f_i)
^{+0.59}_{-0.33}(B_i)
\times 10^{-6}$ \\
& $1.29^{+0.14}_{-0.14}(\beta_{B_c})
^{+0.33}_{-0.30}(f_i)
^{+0.44}_{-0.44}(B_i)
\times 10^{-6}$ \\
\hline
\multirow{2}{*}{$B_c^+ \to f_0(1500) a_2(1320)^+$}
& $0.42^{+0.05}_{-0.04}(\beta_{B_c})
^{+0.11}_{-0.10}(f_i)
^{+0.27}_{-0.15}(B_i)
\times 10^{-6}$ \\
& $
0.58^{+0.06}_{-0.06}(\beta_{B_c})
^{+0.15}_{-0.14}(f_i)
^{+0.20}_{-0.20}(B_i)
\times 10^{-6}$ \\
\hline
Decay modes ($\Delta S=1$)   & Branching ratios \\
\hline
\multirow{2}{*}{$B_c^+ \to f_0(1370) {K^*_2}(1430)^+$}
& $6.26^{+0.70}_{-0.67}(\beta_{B_c})
^{+1.59}_{-1.43}(f_i)
^{+3.30}_{-1.92}(B_i)
\times 10^{-8}$ \\
& $4.17^{+0.34}_{-0.37}(\beta_{B_c})
^{+0.80}_{-0.74}(f_i)
^{+1.49}_{-1.68}(B_i)
\times 10^{-8}$ \\
\hline
\multirow{2}{*}{$B_c^+ \to f_0(1500) {K^*_2}(1430)^+$}
& $4.30^{+0.44}_{-0.43}(\beta_{B_c})
^{+1.06}_{-0.97}(f_i)
^{+1.09}_{-0.73}(B_i)
\times 10^{-8}$ \\
& $6.11^{+0.30}_{-0.25}(\beta_{B_c})
^{+2.23}_{-3.55}(f_i)
^{+0.23}_{-2.40}(B_i)
\times 10^{-8}$ \\
\hline
\end{tabular}}
\end{center}
\end{table}

Notice that, for the $B_c^+ \to f_0(1370, 1500) a_2^+$ modes, the branching ratios in two scenarios
are compatible to each other within still large uncertainties, i.e., ${\rm BR}(B_c^+ \to f_0(1370) a_2^+)_{S1}
=(0.93^{+0.65}_{-0.40})\times 10^{-6} \sim {\rm BR}(B_c^+ \to f_0(1370) a_2^+)_{S2}
=(1.29^{+0.57}_{-0.55})\times 10^{-6}$ and ${\rm BR}(B_c^+ \to f_0(1500) a_2^+)_{S1}
=(0.42^{+0.29}_{-0.18})\times 10^{-6} \sim {\rm BR}(B_c^+ \to f_0(1500) a_2^+)_{S2}
=(0.58^{+0.26}_{-0.25})\times 10^{-6}$.
It seems hard to differentiate the preferred scenario based on the current results,
even when the related measurements could be available
in the future. However, at least, the experimental confirmations on these values could provide evidences
to support the viewpoint that $f_0(1500)$ may not be considered as a primary scalar glueball.
Simultaneously, the $f_0(1370)$ is predominated by the quark-flavor nonstrange components.
Of course, the results with enough small errors in both experiment and theory would help find out
the preferred scenario of $f_0(1370, 1500)$ as ground scalar states or the first-excited ones.

\begin{figure}[!htb]
    \centering
    \includegraphics[width=0.45\textwidth]{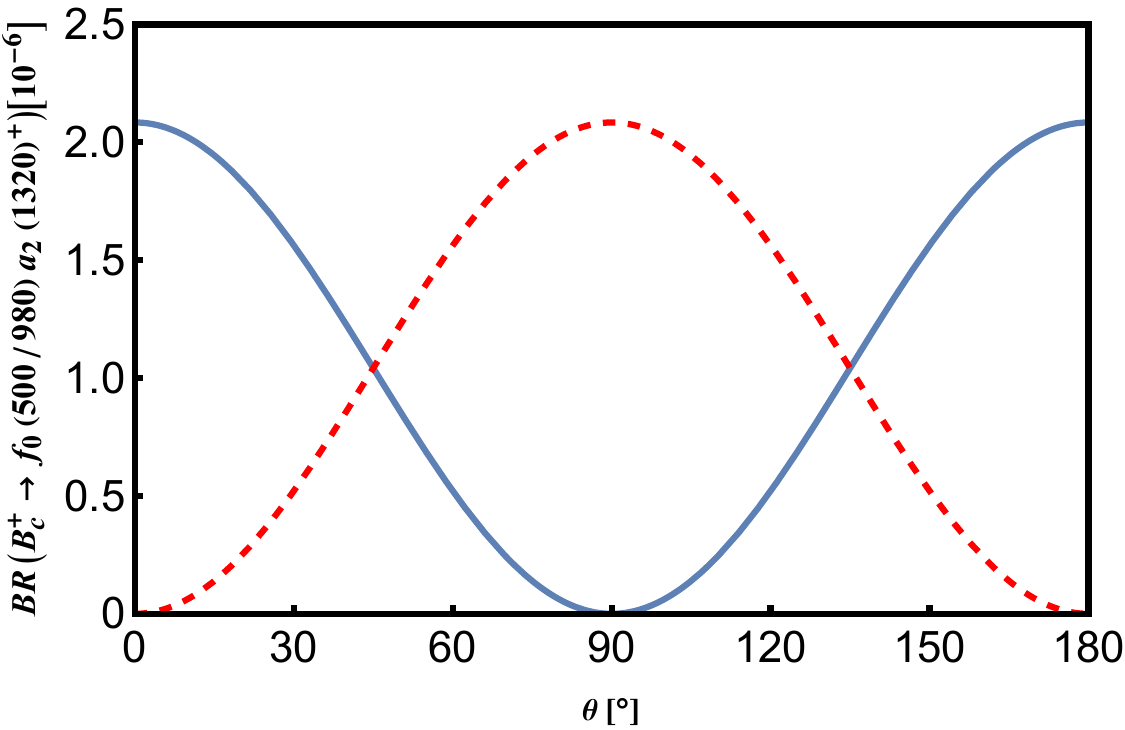}
    \vspace{-0.15cm}
    \caption{Dependence of branching ratios on $\theta$: the solid (dashed) line corresponds to
    ${\rm BR}(B_c^+ \to f_0(500)[f_0(980)] a_2(1320)^+)$.    }
    \label{fig:A}
\end{figure}

\begin{figure}[!ht]
    \centering
    \includegraphics[width=0.45\textwidth]{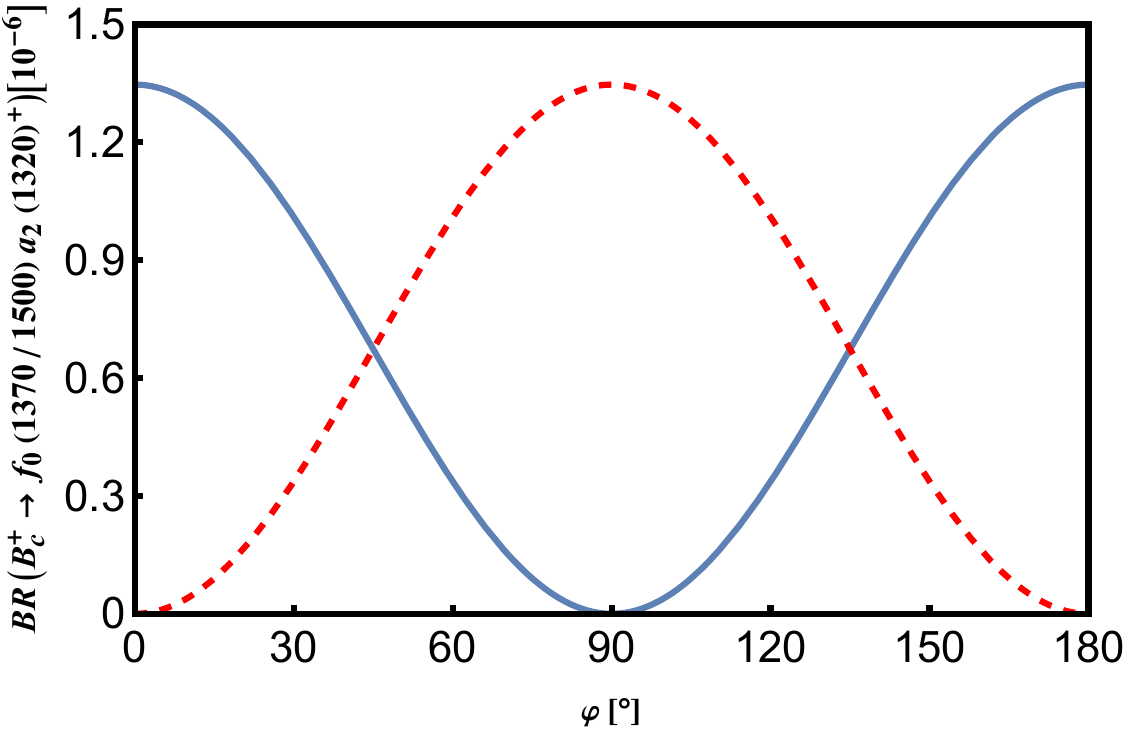}
    \vspace{-0.15cm}
    \caption{Dependence of branching ratios on $\varphi$: the solid (dashed) line corresponds to ${\rm BR}(B_c^+ \to f_0(1370)[f_0(1500)] a_2(1320)^+)$ in $S1$.}
    \label{fig:B}
\end{figure}
\begin{figure}[!ht]
    \centering
    \includegraphics[width=0.45\textwidth]{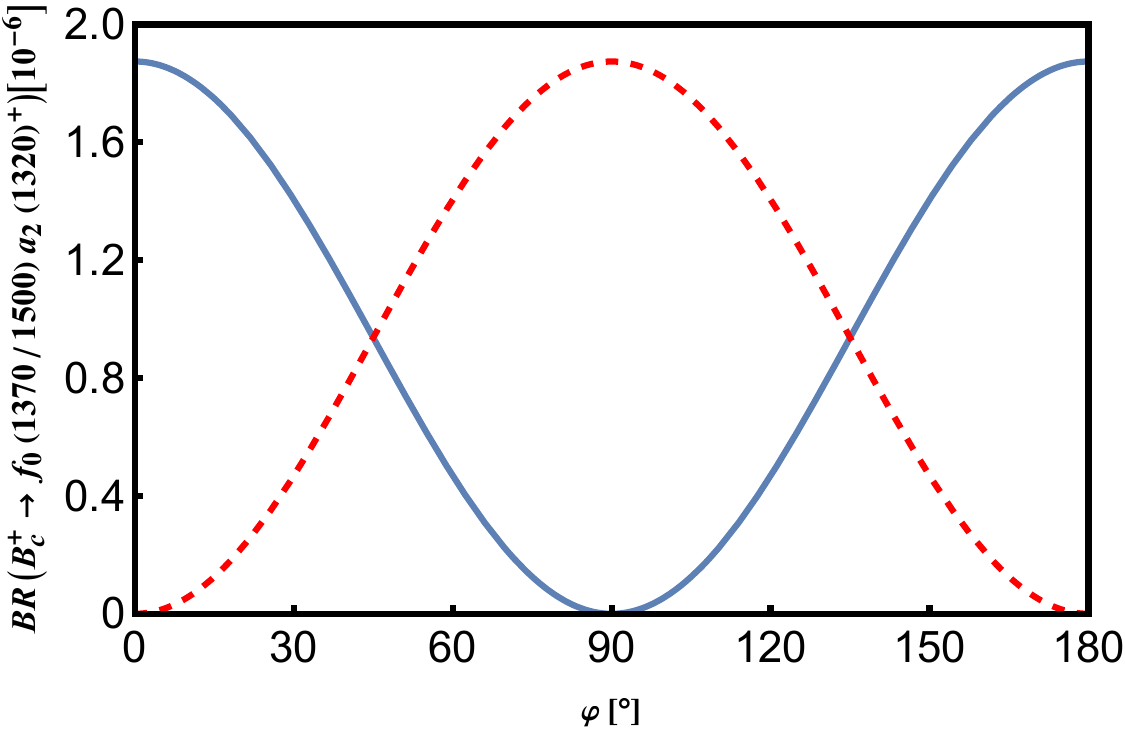}
    \vspace{-0.15cm}
    \caption{Dependence of branching ratios on $\varphi$: the solid (dashed) line corresponds to ${\rm BR}(B_c^+ \to f_0(1370)[f_0(1500)] a_2(1320)^+)$ in $S2$.}
    \label{fig:C}
\end{figure}

To further explore the information of the mixing angles $\theta$ for $f_0(500)-f_0(980)$ and $\varphi$
for $f_0(1370)-f_0(1500)$ through the decays of $B_c^+ \to f_0 a_2^+$ and $f_0^\prime a_2^+$,
we plot their branching ratios, which are naively proportional to the coefficients $\cos^2\theta [\cos^2\varphi]$
and $\sin^2\theta [\sin^2\varphi]$, respectively.
These coefficients vary with the mixing angles in the region of $\theta, \varphi \in [0^\circ, 180^\circ]$
in Figs.~\ref{fig:A}-\ref{fig:C}. In principle, due to the contributions coming from both nonstrange and strange
components of $f_0(500, 980, 1370, 1500)$ through constructive or destructive interferences, the decays of $B_c^+ \to f_0 K_2^{*+}$ and $B_c^+ \to f_0^\prime K_2^{*+}$
can also provide related information to help constrain the mixing angles $\theta$ and $\varphi$. However,
because of the CKM-suppressed factor around 20, the branching ratios are thus too significantly small to be probed 
optimistically in the near future experiments.

%
\section{Conclusions and Summary}
\label{sect:C&S}
In summary, we have investigated the $B_c^+ \to ST$ decay for the first time
within the self-consistent PQCD factorization framework, i.e., the iPQCD formalism,
at leading order in the standard model. Due to the unclear inner structure of the light scalar
states, especially those below 1~GeV, we adopt the assumption of two-quark structure
for light scalars to make the related calculations in the iPQCD formalism. Based on
the two-scenarios classification for the light scalar mesons below 1~GeV and around
1.5~GeV, we obtain the branching ratios in association with relative ratios for the
CKM-favored and CKM-suppressed $B_c^+ \to ST$ modes. By employing the numerical results
computed in the iPQCD formalism, we find that
\begin{itemize}
\item[]{(a)}
the CKM enhanced factor of $|V_{ud}/V_{us}|^2$ near 20 results in large
branching ratios for most $\Delta S=0$ processes generally, except for several modes with significant
destructive interferences, e.g., $B_c^+ \to a_0(980) a_2(1320)$.

\item[]{(b)}
the large branching ratios of approximately $10^{-6}$ predicted in the iPQCD formalism are accessible
at LHCb detector and would be tested in the near-future experiments;

\item[]{(c)}
the verification of these iPQCD branching ratios and the typical relative ratios could provide
opportunities for exploring the nature of the mentioned light scalars, as well as
for testing the reliability of the employed iPQCD formalism;

\item[]{(d)}
the theoretical uncertainties arising from hadronic parameters, such as the
scalar decay constant and Gegenbauer moments of light scalar mesons, remain
evidently large. Undoubtedly, these uncertainties require further improvements
from both lattice QCD calculations
and high-precision experimental measurements.

\end{itemize}


\begin{acknowledgments}
S.W. thanks Y.S. for helpful discussions.
This work is supported by the National Natural Science Foundation of China under the
Grants No.~11875033 and No.~12335003, and
by the Project of Jiangsu Normal University (JSNUGZL2026087).
S.W. is supported by Postgraduate Research $\&$ Practice Innovation Program of Jiangsu Normal
University (2024XKT2334).
\end{acknowledgments}

\medskip
\noindent{\bf Data Availability Statement}. Correspondence and requests for data should
be addressed to S.W. or X.L.

\medskip
\noindent{\bf Declaration of Competing Interest}.
The authors declare no competing interests.

%
%

\begin{appendix}
	
\section{Distribution amplitudes}
\label{a1}

As discussed in~\cite{Liu:2018kuo}, the distribution amplitude $\phi_{B_c}(x, \boldsymbol{b})$
in the impact $\boldsymbol{b}$ space could be read as,
\beq
\phi_{B_c}(x, \boldsymbol{b}) =   \frac{f_{B_c}}{2\sqrt{2 N_c } }
N_{B_c}x(1-x)\exp\left[-\frac{(1-x)m_c^2+xm_b^2}
{8\beta_{B_c}^2x(1-x)}\right]\exp\left[-2\beta_{B_c}^2x(1-x) \boldsymbol{b}^2\right]\;,
\label{eq:a1}
\eeq
where the decay constant $f_{B_c} = 0.489 \pm 0.005$~\cite{Chiu:2007km}. As shown in~\cite{Liu:2023kxr,Liu:2025ipe},
the shape parameter $\beta_{B_c}=1.0 \pm 0.1$ GeV is currently the preferred value in phenomenology.
The $N_{B_c}$
is a normalization constant that can be determined by using the following relation
\beq
\int_0^1 \phi_{B_c}(x, \boldsymbol{b}=0) {\rm d}x \equiv \int_0^1 \phi_{B_c}(x) {\rm d}x
&=& \frac{f_{B_c}}{2\sqrt{2 N_c} }\;.
\label{eq:a2}
\eeq

For the light scalar mesons, the leading-twist distribution amplitude is parameterized as~\cite{Cheng:2005nb,Cheng:2007st,Li:2008tk},
\beq
\phi_S(x) & = & \frac{3}{\sqrt{2N_c}} x (1-x) \left[f_{S}(\mu) + \bar{f}_{S}(\mu) \sum_{m=1}^{\infty}B_{m}(\mu) C_{m}^{3/2} (2x-1) \right] \;,
\label{eq:a3}
\eeq
where $f_{S}(\mu)$ and $\bar{f}_{S}(\mu)$, $B_{m}(\mu)$, and $C_{m}^{3/2}(t)$ are the vector and scalar decay constants,
Gegenbauer moments, and Gegenbauer polynomials, respectively.
The vector and scalar decay constants of a scalar meson are defined by
\beq
\left \langle S(P)\left | \bar{q}_2 \gamma_\mu q_1  \right | 0 \right \rangle &=& f_S P_\mu\;,
\qquad
\left \langle S\left | \bar{q}_2 q_1  \right | 0 \right \rangle = m_S \bar{f}_S\;,
\eeq
An interesting relationship exits between them as follows,
\beq
\bar{f}_{S} &=& \mu_{S} f_{S} \;,
\qquad
\mu_{S} = \frac{m_{S}}{m_{2}(\mu) - m_{1}(\mu)}\;,
\eeq
where $m_{1}$ and $m_{2}$ are the current-quark masses of scalar meson with the running scale $\mu$.
For the neutral scalar mesons $a_{0}^{0}$, $f_{0}$, and $f_0^\prime$, the coupling $f_{\!S}$ vanishes.

According to the available literature, the scalar decay
constants and Gegenbauer moments at the renormalization scale $\mu =1$~GeV in the
LCDAs of quark-flavor scalar states $f_0^n$ and $f_0^s$ are~\cite{Cheng:2005ye},
\beq
\bar f_{f_0^n} &\simeq& 0.35 \;,
\qquad
B_1^n = -0.92\pm 0.08 \;,
\qquad
B_3^n = -1.00\pm 0.05 \;,
\\
\bar f_{f_0^s} &\simeq& 0.33 \;,
\qquad
B_{1,3}^s \simeq 0.8 B_{1,3}^n\;.
\eeq
and those for the quark-flavor scalar states $f_0^N$ and $f_0^S$ are~\cite{Ren:2023ebq},
\beq
\bar f_{f_0^N} &=& \biggl\{\begin{array}{cc}
-0.280\pm 0.030 \qquad (S1)
\\
\hspace{3mm}
0.460\pm 0.050 \qquad (S2)
\end{array} \;,
\qquad
\bar f_{f_0^S} = \biggl\{\begin{array}{cc}
-0.255\pm 0.030 \qquad (S1)
\\
\hspace{3mm}
0.490\pm 0.050 \qquad  (S2)
\end{array}\;,
\\
B_1^{N} &=& \biggl\{\begin{array}{cc}
\hspace{3mm}
1.00 \pm 0.50 \quad (S1)
\\
-0.60\pm 0.14 \quad (S2)
\end{array} \;,
\quad
B_3^{N} = \biggl\{\begin{array}{cc}
-1.65\pm 0.18 \quad (S1)
\\
-0.46\pm 0.25 \quad (S2)
\end{array} \;,
\quad
B_{1,3}^{S}  = 0.8 B_{1,3}^{N}\;.
\eeq
And the decay constants and Gegenbauer moments of the rest scalar states
such as $a_0$ and $K_0^*$ could refer to Ref.~\cite{Cheng:2005nb} for detail
and have been collected in Table~\ref{t:a1}.

\begin{table}[hbt]
\caption{ Decay constants (in units of GeV) and Gegenbauer moments
for light scalars (at $\mu =1$ GeV) from Ref.~\cite{Cheng:2005nb}.}
\label{t:a1}
\begin{center}\vspace{-0.10cm} {\footnotesize
\begin{tabular}[t]{c|c|c|c}
\hline\hline
Scalars & $\bar{f_S}$ & $B1$ & $B3$   \\
\hline
$a_0(980)$ & $0.365 \pm 0.020$ & $-0.93 \pm 0.10$ & $0.14 \pm 0.08$
\\ \hline
$K_0^*(700)$ & $0.340 \pm 0.020$ & $-0.92 \pm 0.11$ &$0.15 \pm 0.09$
\\ \hline
\multirow{2}{*}{$a_0(1450)$} & $-0.280 \pm 0.030$ & $0.89 \pm 0.20$ & $-1.38 \pm 0.18$ \\
& $0.460 \pm 0.050$ & $-0.58 \pm 0.12$ & $-0.49 \pm 0.15$
\\ \hline
\multirow{2}{*}{$K_0^*(1430)$} & $-0.300 \pm 0.030$ & $0.58 \pm 0.07$ & $-1.20 \pm 0.08$ \\
& $0.445 \pm 0.050$ & $-0.57 \pm 0.13$ & $-0.42 \pm 0.22$
\\
\hline \hline
\end{tabular}
}
\end{center}
\end{table}

For the twist-3 distribution amplitudes of scalar meson, the following asymptotic forms are
adopted,
\beq
\phi_{S}^S(x) &=& \frac{1}{2\sqrt{2N_c}}\bar{f}_{S}\;,
\qquad
\phi_{S}^T(x) = \frac{1}{2\sqrt{2N_c}}\bar{f}_{S}(1-2x)\;.
\eeq

For the light tensor mesons, due to the conservation of angular momentum in decay processes,
only the longitudinal LCDA contributes to the decay amplitude.
However, owing to the limited knowledge regarding the LCDAs of tensor mesons,
the available LCDAs are adopted as follows~\cite{Cheng:2010hn,Wang:2010ni},
\beq
\phi_{T}(x) &=& \frac{3f_{T}}{\sqrt{2N_c}} \phi_\parallel(x)\;,
\qquad
\phi_{T}^t(x) = \frac{f_{T}^T}{2\sqrt{2N_c}} h_\parallel^t(x)\;,
\qquad
\phi_{T}^s(x) = \frac{f_{T}^T}{4\sqrt{2N_c}} \frac{d}{dx}h_\parallel^s(x)\;,
\label{eq:a7}
\eeq
with $f_T$ and $f_T^T$ being the vector and tensor decay constants, and
\beq
\phi_\parallel(x)&=& x(1-x)[a_1 C_1^{3/2}(t)]\;,
\label{eq:twist-2}
\quad
h_\parallel^t(x)= \frac{15}{2}(1-6x+6x^2)t\;,
\quad
h_\parallel^s(x)= 15x(1-x)t \;.
\eeq
in which, $a_1=5/3$ and $t=2x-1$.
The decay constants of light tensor mesons are collected in Table~\ref{t:a2}.

\begin{table}[hbt]
\caption{ Decay constants (in units of GeV) for light tensors from Ref.~\cite{Cheng:2010hn}.}
\label{t:a2}
\begin{center}\vspace{-0.10cm} {\footnotesize
\begin{tabular}{cccccccccc}
\hline\hline
   $f_{a_2} $   & $ f_{a_2}^T $ & $ f_{K_2^*} $ & $ f_{K_2^*}^T $  & $f_{f_2} $ & $f_{f_2}^T$ & $f_{f_2^{\prime}} $ & $f_{f_2^{\prime}}^T$ \\
   $0.107\pm0.006$ & $0.105\pm 0.021$ & $ 0.118\pm 0.005$  & $0.077\pm 0.014$ & $0.102\pm 0.006$ & $0.117\pm0.025$    & $0.126\pm 0.004$  & $0.065\pm 0.012$\\
\hline \hline
\end{tabular}}
\end{center}
\end{table}

\section{Factorization formulas and Decay amplitudes}
\label{a3}
In this section, the decay amplitudes of the decay $B_c^+ \to ST$ are given explicitly,
associated with the factorization formulas calculated in the iPQCD formalism.
\begin{itemize}
\item{$B_c^+ \to a_0 T$}
\beq
\sqrt{2}\cdot{\cal A}(B_c^+\to a_0^+ a_2^0)&=& V_{cb}^* V_{ud}\left\{a_1[F_{a_0^+ a_2^u}
- F_{a_2^d a_0^+}] +C_1 [M_{a_0^+a_2^u} - M_{a_2^d a_0^+}]\right\}\;,
\\
\sqrt{2}\cdot{\cal A}(B_c^+\to a_2^+a_0^0)&=&V_{cb}^* V_{ud}\left\{a_1[F_{a_2^u a_0^+} - F_{a_0^+ a_2^d}]
+C_1[M_{a_2^u a_0^+} - M_{a_0^+ a_2^d}]\right\}\;,
\\
{\cal A}(B_c^+\to a_0^+ f_2)&=& V_{cb}^* V_{ud}\left\{a_1[F_{a_0^+ f_2^u}+ F_{f_2^d a_0^+}]
+C_1 [ M_{a_0^+f_2^u} + M_{f_2^d a_0^+}]\right\}\;,
\\
{\cal A}\left(B_c^+\to K_2^{*0}a_0^+\right)&=&V_{cb}^*V_{us} \left\{a_1 F_{K_2^{*0} a_0^+} + C_1 M_{K_2^{*0} a_0^+}\right\}\;,
\\
\sqrt{2}\cdot{\cal A}\left(B_c^+\to K_2^{*+} a_0^0\right)&=&V_{cb}^*V_{us}\left\{a_1F_{K_2^{*+} a_0^0}+C_1 M_{K_2^{*+} a_0^0}\right\}\;,
\eeq

\item{$B_c^+ \to K_0^* T$}
\beq
{\cal A}\left(B_c^+ \to \bar{K}_0^{*0} K_2^{*+}\right)&=&
V_{cb}^*V_{ud}\left\{a_1 F_{K_0^{*0}K_2^{*+}} +C_1 M_{K_0^{*0}K_2^{*+}}\right\}\;,
\\
{\cal A}\left(B_c^+ \to \bar{K}_2^{*0} K_0^{*+}\right)&=&
V_{cb}^*V_{ud}\left\{a_1 F_{K_2^{*0}K_0^{*+}} + C_1 M_{K_2^{*0} K_0^{*+}} \right\}\;,
\\
{\cal A}\left(B_c^+ \to \bar{K}_0^{*0} a_2^+\right)&=&
V_{cb}^*V_{us}\left\{a_1 F_{K_0^{*0}a_2^+} + C_1 M_{K_0^{*0}a_2^+}\right\}\;,
\\
\sqrt{2}\cdot{\cal A}\left(B_c^+\to K_0^{*+} a_2^0\right)&=&
V_{cb}^*V_{us}\left\{a_1F_{K_0^{*+}a_2^0} + C_1 M_{K_0^{*+}a_2^0}\right\}\;,
\\
{\cal A}\left(B_c^+\to K_0^{*+} f_2\right)&=&
V_{cb}^*V_{us}\left\{a_1F_{K_0^{*+}f_2^u}+ C_1 M_{K_0^{*+}f_2^u}\right\}\;,
\\
{\cal A}\left(B_c^+\to K_0^{*+} f_2^{\prime}\right)&=&
V_{cb}^*V_{us}\left\{a_1F_{f_2^\prime K_0^{*+}} + C_1 M_{f_2^\prime K_0^{*+}}\right\}\;,
\eeq

\item{$B_c^+ \to (f_0, f_0^\prime) T$}
\beq
{\cal A}(B_c^+\to a_2^+ f_0(500)) &=& V_{cb}^* V_{ud} \cos{\theta}
\biggl\{  a_1 \left[F_{a_2^+ f_{0}^{n_d}}+ F_{f_{0}^{n_u} a_2^+}\right]
+ C_1\left[ M_{a_2^+ f_{0}^{n_d}}+ M_{f_{0}^{n_u} a_2^+} \right]\biggr\}\;,
\\
{\cal A}(B_c^+\to a_2^+ f_0(980)) &=& V_{cb}^* V_{ud}\sin{\theta}
\biggl\{a_1\left[ F_{a_2^+ f_{0}^{n_d}}+ F_{f_{0}^{n_u} a_2^+}\right]
+ C_1 \left[ M_{a_2^+ f_{0}^{n_d}} + M_{f_{0}^{n_u} a_2^+}\right]\biggr\}\;,
\eeq
\beq
{\cal A}(B_c^+\to K_2^{*+} f_0(500))&=& V_{cb}^*V_{us}
\biggl\{ \cos{\theta}\left[a_1 F_{K_2^{*} f_{0}^{n_u}}+C_1 M_{K_2^{*} f_{0}^{n_u}}\right]
- \sin{\theta}\left[a_1 F_{f_{0}^s K_2^{*}}+C_1 M_{f_{0}^s K_2^{*}}\right]\biggr\}\;,
\\
{\cal A}(B_c^+\to K_2^{*+} f_0(980))&=&V_{cb}^*V_{us}
\biggl\{ \sin{\theta}\left[a_1 F_{K_2^{*} f_{0}^{n_u}}+C_1 M_{K_2^{*} f_{0}^{n_u}}\right]
+\cos{\theta} \left[a_1 F_{f_{0}^s K_2^{*}} +C_1 M_{f_{0}^s K_2^{*}}\right]\biggr\}\;,
\eeq
in which, the decay amplitude ${\cal A}(B_c^+ \to f_0(1370, 1500) T)$ could be easily obtained
by the replacements of $\theta \to \varphi$,
$f_0^n \to f_0^N$, and $f_0^s \to f_0^S$ correspondingly.
\end{itemize}

In the above decay amplitudes, the related factorization formulas are presented as follows,
\beq
F_{ST} &=& 8\sqrt{\frac{2}{3}}\pi C_F f_{B_c} m_{B_c}^2 \int_0^1{\rm d}x_2 {\rm d}x_3
\int_0^{\infty}b_2{\rm d}b_2b_3{\rm d}b_3  \non
& & \times \biggl\{ \biggl[ \left(1- x_3\right)\phi_S(x_2) \phi_T(x_3)
- 2 r_2 r_3 \phi_S^S(x_2) \left(x_3 \phi_T^t(x_3) -(x_3 - 2) \phi_T^s(x_3)
\right)  \biggr]
\non
&& \cdot E_{f}(t_a)S_t(x_3)h(\alpha_{f1},\beta_{f1},b_2,b_3)
\non
& &
-\biggl[  x_2 \phi_S(x_2) \phi_T(x_3) -2 r_2  r_3 \phi_T^s(x_3)
\left((x_2+1)\phi_S^S(x_2) + (x_2 -1)\phi_S^T(x_2) \right)\biggr]
\non
& &  \cdot E_{f}(t_b)S_t(x_2)h(\alpha_{f2},\beta_{f2},b_3,b_2) \biggr\}\;,
\eeq
\beq
M_{ST} &=&  \frac{32}{3}\pi C_F m_{B_c}^2\int_0^1{\rm d}x_1{\rm d}x_2{\rm d}x_3
\int_0^{\infty}b_1{\rm d}b_1b_2{\rm d}b_2 \phi_{B_c}(x_1, b_1)
\non
&& \times \biggl\{ \biggl[ \left( r_b +x_1+x_2-1\right)\phi_S(x_2) \phi_T(x_3)
+ r_2r_3\left(\phi_S^S(x_2) \left( \left(1-2x_1-x_2+x_3-4r_b\right)\phi_T^s(x_3)\right.\right.
\non && \left. \left.
-\left(1-x_2-x_3 \right)\phi_T^t(x_3) \right) -\phi_S^T(x_2) \left( \left(1-2x_1-x_2+x_3\right)\phi_T^t(x_3)
+  \left(1-x_2-x_3\right)\phi_T^s(x_3)\right)\right)  \biggr]
 \non
&& \cdot E_{nf}(t_c)h(\alpha_{nf1},\beta_{nf1},b_1,b_2)
\non
&-&
\biggl[\left(1-x_1-x_3+r_c\right)\phi_S(x_2)\phi_T(x_3)+ r_2r_3\left(\phi_S^S(x_2)\left(\left(1-x_2-x_3\right)\phi_T^t(x_3)-\left(1-2x_1+x_2-x_3
\right.\right.\right.
\non
&& \left.\left.\left.
+4r_c\right)\phi_T^s(x_3) \right)+ \phi_S^T(x_2)\left(\left(1-2x_1+x_2-x_3\right)\phi^t_T(x_3)-\left(1-x_2-x_3\right)\phi_T^s(x_3)
\right)\right)  \biggr]
\non &&
\cdot E_{nf}(t_d)h(\alpha_{nf2},\beta_{nf2},b_1,b_2) \biggr\} \;,
\eeq
\beq
F_{TS} &=& 8\sqrt{\frac{2}{3}}\pi C_F f_{B_c} m_{B_c}^2
\int_0^1{\rm d}x_2{\rm d}x_3\int_0^{\infty}b_2{\rm d}b_2b_3{\rm d}b_3
\non
& & \times \biggl\{ \biggl[ \left(1- x_3\right) \phi_T(x_2) \phi_S(x_3) - 2r_2r_3 \phi_T^s(x_2)
\left( x_3 \phi_S^T(x_3)-\left(x_3-2\right)\phi_S^S(x_3)\right) \biggr]
\non &&
\cdot
E_{f}(t_a)S_t(x_3)h(\alpha_{f1},\beta_{f1},b_2,b_3)
\non
& & +\biggl[x_2 \phi_T(x_2)\phi_S(x_3)- 2r_2r_3\phi_S^S(x_3)\left(\left(x_2+1\right)\phi_T^s(x_2)+ \left(x_2-1\right)\phi_T^t(x_2)\right) \biggr]
\non &&
\cdot
E_{f}(t_b)S_t(x_2)h(\alpha_{f2},\beta_{f2},b_3,b_2)\biggr\}
\eeq
\beq
M_{TS} &=&  \frac{32}{3}\pi C_F m_{B_c}^2\int_0^1{\rm d}x_1{\rm d}x_2{\rm d}x_3
\int_0^{\infty}b_1{\rm d}b_1b_2{\rm d}b_2  \phi_{B_c}(x_1, b_1)
\non
& & \times  \biggl\{ \biggl[\left(r_b+x_1+x_2-1\right)\phi_T(x_2)\phi_S(x_3)
+
r_2r_3\left(\phi_T^s(x_2)\left(\left(1-2x_1-x_2+x_3-4r_b\right)\phi_S^S(x_3)\right.\right.
\non
& & \left.\left.
-\left(1-x_2-x_3\right)\phi_S^T(x_3)\right)+ \phi_T^t(x_2)\left(\left(1-x_2-x_3\right)\phi_S^S(x_3)
-\left(1-2x_1-x_2+x_3\right)\phi_S^T(x_3) \right)\right) \biggr]
\non
&& \cdot E_{nf}(t_c)h(\alpha_{nf1},\beta_{nf1},b_1,b_2)
\non
& &   -\biggl[\left(1-x_1-x_3+r_c\right)\phi_T(x_2) \phi_S(x_3) +r_2 r_3
\left(\phi_T^s(x_2)\left(\left(1-x_2-x_3\right)\phi_S^T(x_3) -\left(1-2x_1+x_2-x_3 \right.\right.\right.
\non
&& \left.\left.\left.
+4r_c\right)\phi_S^S(x_3) \right)+\phi_T^t(x_2)\left(\left(1-2x_1+x_2-x_3\right)\phi_S^T(x_3) -\left(1-x_2-x_3\right)\phi_S^S(x_3)\right)\right)\biggr]
\non
&&
\cdot
E_{nf}(t_d)h(\alpha_{nf2},\beta_{nf2},b_1,b_2)\biggr\}
\eeq
where the hard function can be read as,
\beq
h(\alpha,\beta,b_i,b_j)=K_0(\beta b_j)\big[\theta(b_j-b_i)I_0(\alpha b_i)K_0(\alpha b_j)+\theta(b_i-b_j)I_0(\alpha b_j)K_0(\alpha b_i) \big]
\eeq
with the parameters $\alpha$ and $\beta$
\beq
\alpha_{f1}^2&=&m_{B_c}^2(1+(r_2^2-1)x_3)(r_3^2x_3-1)\;, \non
\beta_{f1}^2&=&m_{B_c}^2(1-x_3+r_2^2(x_2x_3-1))(-x_2+r_3^2(x_2+x_3-1))\;, \non
\alpha_{f2}^2&=&m_{B_c}^2(1+r_2^2(x_2-1))(r_3^2(x_2-1)-x_2)\;, \non
\beta_{f2}^2&=&m_{B_c}^2(1-x_3+r_2^2(x_2+x_3-1))(-x_2+r_3^2(x_2+x_3-1)) \;, \non
\alpha_{nf1}^2 &=&m_{B_c}^2(1-x_3+r_2^2(-1+x_2+x_3))(-x_2+r_3^2(-1+x_2+x_3))\;, \non
\beta_{nf1}^2 &=&m_{B_c}^2r_b^2-m_{B_c}^2(x_1-x_3-r_2^2(1-x_2-x_3))(-1+x_1+x_2+r_3^2(1-x_2-x_3))\;, \non
\alpha_{nf2}^2 &=&m_{B_c}^2(1-x_3-r_2^2(1-x_2-x_3))(-x_2-r_3^2(1-x_2-x_3))\;, \non
\beta_{nf2}^2 &=&m_{B_c}^2r_c^2+m_{B_c}^2(1-x_1-x_3-r_2^2(1-x_2-x_3))(x_1-x_2-r_3^2(1-x_2-x_3))
\eeq
And, if $\alpha^2<0$ or $\beta^2<0$, then the following relations will be utilized,
\beq
I_0(ix)=J_0(x) \hspace{1cm} K_0(-ix)=-\frac{\pi}{2}Y_0(x)+i\frac{\pi}{2}J_0(x).
\eeq

The expressions of evolution functions $E_f(t_i)$ and $E_{nf}(t_i)$ are shown as follows,
\beq
E_f(t_i)=\alpha_s(t_i)S_{ab}(t_i)S_{t}(x) \hspace{1cm} E_{nf}(t_i)=\alpha_s(t_i)S_{cd}(t_i),
\eeq
in which, $\alpha_s(t)$ is the QCD running coulping constant with the hard scales $t_{a,b,c,d}$
\beq
t_a &=& \max(|\sqrt{\alpha_{f1}^2}|,|\sqrt{\beta_{f1}^2}|,1/b_2,1/b_3)\;,
\qquad
t_b = \max(|\sqrt{\alpha_{f2}^2}|,|\sqrt{\beta_{f2}^2}|,1/b_2,1/b_3)\;, \non t_c&=&\max(|\sqrt{\alpha_{nf1}^2}|,|\sqrt{\beta_{nf1}^2}|,1/b_1,1/b_2)\;,
\qquad
t_d =\max(|\sqrt{\alpha_{nf2}^2}|,|\sqrt{\beta_{nf2}^2}|,1/b_1,1/b_2)\;,
\eeq
Moreover, the Sudakov factors are given as,
\beq
S_{ab}(t_i)
&=& s\left(x_2
P_2^+, b_2\right) +s\left((1-x_2) P_2^+,
b_2\right)+s\left(x_3
P_3^-, b_3\right) +s\left((1-x_3) P_3^-,
b_3\right) \non
&& -\frac{1}{\beta_1}\left[\ln\frac{\ln(t_i/\Lambda)}{-\ln(b_2/\Lambda)}
\right]-\frac{1}{\beta_1}\left[\ln\frac{\ln(t_i/\Lambda)}{-\ln(b_3/\Lambda)}
\right],
\\
S_{cd}(t_i) &=& s_c\left(x_1 P_1^+, b_1\right)
 +s\left(x_2 P_2^+, b_2\right)
+s\left((1-x_2) P_2^+, b_2\right) +s\left(x_3
P_3^-, b_1\right) \non
 && +s\left((1-x_3) P_3^-,
b_1\right) -\frac{1}{\beta_1}\left[
\frac{5}{6} \ln\frac{\ln(t_i/\Lambda)}{\ln(m_c/\Lambda)}
+2\ln\frac{\ln(t_i/\Lambda)}{-\ln(b_2\Lambda)}\right]\;,
\eeq
where the explicit expression of the Sudakov exponent $s(Q,b)$ for an energetic light
quark is referred to Ref.~\cite{Li:2003yj} and $s_c(Q,b)$ for the energetic charm quark can be
found in~\cite{Liu:2023kxr}.
And, the jet function $S_t(x)$, arising from threshold resummation, systematically resums
the large double logarithms $\ln^2 x$ that appear in the hard scattering kernels
when the parton momentum fraction approaches to the endpoint region. This function is universal
and has been parameterized in a simplified form, which is independent of the specific decay channel,
twist structure, or quark flavor involved.
\beq
S_t(x)
=\frac{2^{1+2c}\,\Gamma\bigl(\tfrac{3}{2}+c\bigr)}%
{\sqrt{\pi}\,\Gamma(1+c)}\,[x(1-x)]^c \,,
\label{B8}
\eeq
Here, $c = 0.3$ is a phenomenological parameter and also widely adopted in the available literature.
In the non-factorizable decay amplitude, the numerical impact of $S_t(x)$
is usually found to be negligible~\cite{Li:2001ay,Li:2002mi}.

\end{appendix}

%
%


\begin{thebibliography}{99}

\bibitem{CDF:1998axz}
F.~Abe \textit{et al.} [CDF],
Phys. Rev. D \textbf{58}, 112004 (1998)

\bibitem{CDF:1998ihx}
F.~Abe \textit{et al.} [CDF],
Phys. Rev. Lett. \textbf{81}, 2432-2437 (1998)

\bibitem{E791:2000vek}
E.~M.~Aitala \textit{et al.} [E791],
Phys. Rev. Lett. \textbf{86}, 770-774 (2001)

\bibitem{Belle:2002ryx}
K.~Abe \textit{et al.} [Belle],
Phys. Rev. D \textbf{65}, 092005 (2002).

\bibitem{BESIII:2023hml}
M.~Ablikim \textit{et al.} [BESIII],
Phys. Rev. Lett. \textbf{132}, 151903 (2024).

\bibitem{Rosner:2003ia}
J.~L.~Rosner,
Phys. Rev. D \textbf{69}, 094014 (2004).

\bibitem{Vagnoni:2025qfv}
V.~Vagnoni [LHCb],
[arXiv:2503.23087 [hep-ex]].



\bibitem{Cheng:2005nb}
H.~Y.~Cheng, C.~K.~Chua and K.~C.~Yang,
Phys. Rev. D \textbf{73}, 014017 (2006).

\bibitem{Cheng:2007st}
H.~Y.~Cheng, C.~K.~Chua and K.~C.~Yang,
Phys. Rev. D \textbf{77}, 014034 (2008).

\bibitem{Cheng:2013fba}
H.~Y.~Cheng, C.~K.~Chua, K.~C.~Yang and Z.~Q.~Zhang,
Phys. Rev. D \textbf{87}, 114001 (2013).

\bibitem{Wang:2006ria}
W.~Wang, Y.~L.~Shen, Y.~Li and C.~D.~L\"u,
Phys. Rev. D \textbf{74}, 114010 (2006).

\bibitem{Shen:2006ms}
Y.~L.~Shen, W.~Wang, J.~Zhu and C.~D.~L\"u,
Eur. Phys. J. C \textbf{50}, 877 (2007).

\bibitem{Dou:2015mka}
D.~Dou, X.~Liu, J.~W.~Li and Z.~J.~Xiao,
J. Phys. G \textbf{43}, 045001 (2016).

\bibitem{Zou:2016yhb}
Z.~T.~Zou, Y.~Li and X.~Liu,
Phys. Rev. D \textbf{95}, 016011 (2017).

\bibitem{Zou:2017iau}
Z.~T.~Zou, Y.~Li and X.~Liu,
Eur. Phys. J. C \textbf{77}, 870 (2017).


\bibitem{Liu:2019ymi}
X.~Liu, Z.~T.~Zou, Y.~Li and Z.~J.~Xiao,
Phys. Rev. D \textbf{100}, 013006 (2019).

\bibitem{Su:2019vbu}
L.~Su, Z.~Jiang and X.~Liu,
J. Phys. G \textbf{46}, 085003 (2019).

\bibitem{Liu:2021gnp}
Z.~W.~Liu, Z.~T.~Zou, Y.~Li, X.~Liu and J.~Wang,
Eur. Phys. J. C \textbf{82}, 59 (2022).

\bibitem{Niu:2021pcx}
H.~D.~Niu, G.~D.~Li, J.~L.~Ren and X.~Liu,
Eur. Phys. J. C \textbf{82}, 177 (2022).

\bibitem{Li:2025ddc}
M.Q.~Li, X.~Liu, Z.T.~Zou, Y.~Li and Z.J.~Xiao,
Eur. Phys. J. C \textbf{85}, 1300 (2025).



\bibitem{Liu:2010kq}
X.~Liu and Z.~J.~Xiao,
Phys. Rev. D \textbf{82}, 054029 (2010).

\bibitem{Zou:2017yxc}
Z.~T.~Zou, Y.~Li and X.~Liu,
Phys. Rev. D \textbf{97}, 053005 (2018).


\bibitem{Liu:2023kxr}
X.~Liu,
Phys. Rev. D \textbf{108}, 096006 (2023)

\bibitem{Zhang:2025yvt}
W.~J.~Zhang and X.~Liu,
Phys. Rev. D \textbf{113}, 056023 (2026).

\bibitem{Beneke:1999br}
M.~Beneke, G.~Buchalla, M.~Neubert and C.~T.~Sachrajda,
Phys. Rev. Lett. \textbf{83}, 1914 (1999).


\bibitem{Keum:2000wi}
Y.Y.~Keum, H.-n.~Li, and A.I.~Sanda,
Phys.\ Rev.\ D {\bf 63}, 054008 (2001).

\bibitem{Lu:2000em}
C.D.~L\"u, K.~Ukai and M.~Z.~Yang,
Phys. Rev. D \textbf{63}, 074009 (2001).

\bibitem{ParticleDataGroup:2024cfk}
S.~Navas \textit{et al.} [Particle Data Group],
Phys. Rev. D \textbf{110}, 030001 (2024);
``64. Scalar mesons below 1~GeV", mini-review by T.~Gutsche, C.~Hanhart, R.E.~Mitchell and S.~Spanier
in the ``Review of Particle Physics".



\bibitem{Liu:2018kuo}
X.~Liu, H.-n.~Li and Z.~J.~Xiao,
Phys. Rev. D \textbf{97}, 113001 (2018).



\bibitem{Liu:2020upy}
X.~Liu, H.-n.~Li and Z.~J.~Xiao,
Phys. Lett. B \textbf{811}, 135892 (2020).


\bibitem{Liu:2009qa}
X.~Liu, Z.~J.~Xiao and C.~D.~Lu,
Phys. Rev. D \textbf{81}, 014022 (2010).

\bibitem{Liu:2010da}
X.~Liu and Z.~J.~Xiao,
Phys. Rev. D \textbf{81}, 074017 (2010).

\bibitem{Liu:2010epa}
X.~Liu and Z.~J.~Xiao,
J. Phys. G \textbf{38}, 035009 (2011).

\bibitem{Xiao:2011zz}
Z.~J.~Xiao and X.~Liu,
Phys. Rev. D \textbf{84}, 074033 (2011).

\bibitem{Xiao:2013lia}
Z.~J.~Xiao and X.~Liu,
Chin. Sci. Bull. \textbf{59}, 3748 (2014).


\bibitem{Liu:2017cwl}
X.~Liu, R.~H.~Li, Z.~T.~Zou and Z.~J.~Xiao,
Phys. Rev. D \textbf{96}, 013005 (2017).


\bibitem{Liu:2025ipe}
X.~Liu, H.-n.~Li and Z.~J.~Xiao,
Phys. Rev. D \textbf{112}, 036015 (2025).


\bibitem{CMS:2023rev}
A.~Hayrapetyan \textit{et al.} [CMS],
Nature Commun. \textbf{16}, 7990 (2025).

\bibitem{ALICE:2023eyl}
S.~Acharya \textit{et al.} [ALICE],
Phys. Lett. B \textbf{856}, 138915 (2024).



\bibitem{Li:2000zb}
D.~M.~Li, H.~Yu and Q.~X.~Shen,
J. Phys. G \textbf{27}, 807 (2001).

\bibitem{Cheng:2011fk}
H.~Y.~Cheng and R.~Shrock,
Phys. Rev. D \textbf{84}, 094008 (2011).


\bibitem{BESIII:2026mvn}
M.~Ablikim \textit{et al.} [BESIII],
[arXiv:2607.20366 [hep-ex]].

\bibitem{Close:2005vf}
F.~E.~Close and Q.~Zhao,
Phys. Rev. D \textbf{71}, 094022 (2005).

\bibitem{Cheng:2015iaa}
H.~Y.~Cheng, C.~K.~Chua and K.~F.~Liu,
Phys. Rev. D \textbf{92}, 094006 (2015).


\bibitem{Ren:2023ebq}
J.~L.~Ren, M.~Q.~Li, X.~Liu, Z.~T.~Zou, Y.~Li and Z.~J.~Xiao,
Eur. Phys. J. C \textbf{84}, 358 (2024).



\bibitem{Klempt:2021nuf}
E.~Klempt,
Phys. Lett. B \textbf{820}, 136512 (2021).



\bibitem{Stone:2013eaa}
S.~Stone and L.~Zhang,
Phys. Rev. Lett. \textbf{111}, no.6, 062001 (2013).




\bibitem{BESIII:2018ozj}
M.~Ablikim \textit{et al.} [BESIII],
Phys. Rev. Lett. \textbf{121}, 022001 (2018).

\bibitem{Li:2022pkl}
J.~T.~Li, J.~X.~Lin, G.~J.~Zhang, W.~H.~Liang and E.~Oset,
Chin. Phys. C \textbf{46}, 083108 (2022).

\bibitem{Wang:2016wpc}
W.~Wang,
Phys. Lett. B \textbf{759}, 501 (2016).

\bibitem{Roca:2012cv}
L.~Roca,
Phys. Rev. D \textbf{88}, 014045 (2013).

\bibitem{Liu:2010id}
B.~Liu [BESIII],
PoS \textbf{ICHEP2010}, 158 (2010).

\bibitem{BESIII:2010dhc}
M.~Ablikim \textit{et al.} [BESIII],
Phys. Rev. D \textbf{83}, 032003 (2011).

\bibitem{Wu:2008hx}
J.~J.~Wu and B.~S.~Zou,
Phys. Rev. D \textbf{78}, 074017 (2008).



\bibitem{Klempt:2021wpg}
E.~Klempt and A.~V.~Sarantsev,
Phys. Lett. B \textbf{826}, 136906 (2022).





\bibitem{Botts:1989kf}
J.~Botts and G.~F.~Sterman,
Nucl. Phys. B \textbf{325}, 62 (1989).

\bibitem{Li:1992nu}
H.-n.~Li and G.~F.~Sterman,
Nucl. Phys. B \textbf{381}, 129 (1992).

\bibitem{Li:2003yj}
H.-n.~Li,
Prog. Part. Nucl. Phys. \textbf{51}, 85 (2003).

\bibitem{Li:2001ay}
H.-n.~Li,
Phys. Rev. D \textbf{66}, 094010 (2002).

\bibitem{Li:2002mi}
H.-n.~Li and K.~Ukai,
Phys. Lett. B \textbf{555}, 197 (2003).



\bibitem{Li:2008tk}
R.~H.~Li, C.~D.~L\"u, W.~Wang and X.~X.~Wang,
Phys. Rev. D \textbf{79}, 014013 (2009).

\bibitem{Cheng:2010hn}
H.~Y.~Cheng, Y.~Koike and K.~C.~Yang,
Phys. Rev. D \textbf{82}, 054019 (2010).

\bibitem{Wang:2010ni}
W.~Wang,
Phys. Rev. D \textbf{83}, 014008 (2011).


\bibitem{Datta:2007yk}
A.~Datta, Y.~Gao, A.~V.~Gritsan, D.~London, M.~Nagashima and A.~Szynkman,
Phys. Rev. D \textbf{77}, 114025 (2008).



\bibitem{Buchalla:1995vs}
G.~Buchalla, A.~J.~Buras and M.~E.~Lautenbacher,
Rev. Mod. Phys. \textbf{68}, 1125 (1996).



\bibitem{Descotes-Genon:2009eui}
S.~Descotes-Genon, J.~He, E.~Kou and P.~Robbe,
Phys. Rev. D \textbf{80}, 114031 (2009).



\bibitem{Chiu:2007km}
T.~W.~Chiu \textit{et al.} [TWQCD],
Phys. Lett. B \textbf{651}, 171 (2007).


\bibitem{Cheng:2005ye}
H.~Y.~Cheng and K.~C.~Yang,
Phys. Rev. D \textbf{71}, 054020 (2005).






	
\end{thebibliography}
\end{document}